\documentclass[%
 reprint,superscriptaddress,
 amsmath,amssymb, nofootinbib,
 aps,twocolumn,pra]{revtex4-2}

\usepackage{graphicx}
\usepackage{dcolumn}
\usepackage{bm}
\usepackage{comment}
\usepackage{xcolor}
\usepackage{amsmath,amsthm,array}
\usepackage{amssymb,physics,appendix,caption}
\usepackage[colorlinks=true,
linkcolor = blue,
anchorcolor = black,
urlcolor = black,
citecolor = blue
]{hyperref}
\newtheoremstyle{dotless}{}{}{}{}{\bfseries}{}{ }{}
\theoremstyle{dotless}

\newtheorem{chsh}{\textit{CHSH Scenario}}
\newtheorem{prisoner}{\textit{Prisoner Scenario}}
\newtheoremstyle{claim}
  {\topsep}
  {\topsep}
  {}
  {}
  {\itshape}
  {.}
  {.5em}
  {\thmname{#1}\thmnumber{ #2}\thmnote{ (#3)}}
\theoremstyle{claim}
\newtheorem*{uniform}{Uniform}
\newtheorem*{skewC}{Skew Classical}
\newtheorem*{skewQ}{Skew Quantum}

\newtheorem*{lowlow}{Low-Low}
\newtheorem*{highhigh}{High-High}
\newtheorem*{lowhigh}{Low-High}

\usepackage{xpatch}
\makeatletter
\newcommand{\TODO}[1]{\textcolor{red}{\@ifnotempty{#1}{ #1}}}
\makeatother

\usepackage[mathlines]{lineno}

\usepackage[singlelinecheck=false]{subcaption}
\begin{document}


\title{Bayesian rational agents in iterated quantum games}

\author{John B. DeBrota\smallskip}
\affiliation{
Center for Quantum Information and Control, University of New Mexico, Albuquerque, NM 87131, USA
}
\author{Peter J. Love\smallskip}
\affiliation{
Department of Physics and Astronomy, Tufts University,
574 Boston Avenue, Medford MA 02155, USA
}

\begin{abstract}
We apply a Bayesian agent-based framework inspired by QBism to iterations of two quantum games, the CHSH game and the quantum prisoners' dilemma. 
In each two-player game, players hold beliefs about an amount of shared entanglement and about the actions or beliefs of the other player. Each takes actions which maximize their expected utility and revises their beliefs with the classical Bayes rule between rounds. We simulate iterated play to see if and how players can learn about the presence of shared entanglement and to explore how their performance, their beliefs, and the game's structure interrelate. In the CHSH game, we find that players can learn that entanglement is present and use this to achieve quantum advantage. We find that they can only do so if they also believe the other player will act correctly to exploit the entanglement. In the case of low or zero entanglement in the CHSH game, the players cannot achieve quantum advantage, even in the case where they believe the entanglement is higher than it is. For the prisoners dilemma, we show that assuming 1-fold rational players (rational players who believe the other player is also rational) reduces Eisert \emph{et al.}'s quantum extension of the prisoners dilemma to a game with only two strategies, one of which (defect) is dominant for low entanglement, and the other (Eisert \emph{et al.}'s quantum strategy Q) is dominant for high entanglement. For intermediate entanglement, neither strategy is dominant. We again show that players can learn entanglement in iterated play. We also show that strong belief in entanglement causes optimal play even in the absence of entanglement --- showing that belief in entanglement is acting as a proxy for the players trusting each other. Our work may point to future applications in resource detection and quantum algorithm design.

\end{abstract}

\maketitle

\section{Introduction}

Examining games in which the addition of quantum resources allows for a competitive advantage has attracted significant attention~\cite{Clauser1969,Eisert99,Meyer1999,Popescu1994,Popescu2014}. Generally, one seeks games where the addition of quantum resources such as entanglement, contextuality, discord, and/or enlarged sets of possible actions allows for quantum strategies which strictly outperform those available to players with only classical resources. If we can obtain an intuition for finding these gaps in performance, this might inform quantum algorithm design~\cite{Shor2003}. 

Much of the difficulty in finding useful quantum algorithms likely stems from confusion regarding quantum theory itself.\footnote{This is certainly true in the case of the authors!} How we interpret quantum resources is of critical importance to any conclusions drawn from their presence. In the game setting, if we think quantum states, channels, and measurements are part of agent-independent reality, then these objects can appear in a game's description. On the other hand, if we take a broadly epistemic view, these elements will depend, at least in part, on the knowledge and beliefs of the agent who uses them. The perspective one takes will inspire different explanations for the origin of quantum advantage and different directions in which to explore further. This is an opportunity for quantum foundations to directly contribute to quantum algorithm design.

Discussion of quantum games typically does not address foundational questions; when it does, authors usually take the ontic perspective that quantum states, channels, and measurements are part of agent-independent reality. We confront this by explicitly fixing an interpretation in the epistemic tradition which is remarkably well-suited to this context, the quantum Bayesian interpretation, QBism~\cite{Fuchs2013,Fuchs2016a}. 

QBism goes further than any other interpretation in regarding the formal apparatus of quantum theory to concern the knowledge and beliefs of actors, or agents, in the world. Quantum theory, according to this interpretation, is an addition to Bayesian decision theory. Consequently, although the structure of QBism is objectively imposed by nature, the objects of the theory, that is, quantum states, channels, and measurements, are relative to a reasoning subject. To properly use quantum theory, an individual strives for \emph{self} consistency.

This perspective is well-suited to the study of quantum games. Game performance derives from players' strategies, that is, their plans to turn their beliefs into actions. Classical game theory is mostly concerned with game structure, largely glossing over the process of reasoning about one's opponents~\cite{vonNeumann2007}. By contrast, \emph{epistemic} game theory \cite{Pacuit2017,Dekel2015,Perea2012} broadens the scope of the study of games by explicitly including the beliefs of the players. Indeed, in the course of an actual game, an individual's strategy may depend on more than the structure of the game, e.g., their beliefs about the proclivities, capabilities, and beliefs of their opponents. (In \emph{The Dark Forest}, Cixin Liu's ``chain of suspicion'' is an example of this~\cite{Liu2015}.) In general, each player can have different beliefs; for a QBist, this means, among other differences, that the quantum states used by distinct agents in their decision making need not agree. A quantum Bayesian analysis of games constitutes a shift from a global, objective, state-based view, to an agent-based view centered on the collective belief dynamics of individual players. 

Because a strategy is a decision rule, the preferred actions of players are inherited from decision theory. We will work with probability theory which emerges from the classical theory of Bayesian decision making~\cite{Bernardo2000}. In this context, a \emph{rational agent} is one who always chooses to take actions which maximize their expected utility. Following QBism, we note that all of the objects in quantum theory obtain probabilistic representations through an informationally complete reference measurement~\cite{DeBrota2020d}. Thus, a simple model of a user of quantum theory is simply a rational agent whose probabilities satisfy additional constraints. We developed and explored interactions between such agents in~\cite{DeBrota21}. It is then straightforward to apply our framework to make ``quantized'' rational agents interact according to the rules of a game, which is the topic of the present paper.

Bayesian rational behavior does not necessarily coincide with optimal play from a classical game theoretic perspective because it depends on the priors of each agent. To complicate matters, one often sees the term ``rational'' in the classical game theory literature associated with playing a Nash equilibrium, seeming to provide a conception of rationality applicable in the absence of priors. This perspective has received criticism \cite{Risse00,Perea2012,Dekel2015}. Instead, from an epistemic position, Nash equilibrium play arises from the special case of rational reasoning where any given first player believes all of their opponents correctly knows the first's beliefs and also believes their opponents share the first's own beliefs about other players~\cite{Perea2012}. While this situation is possible, it is far from general. To avoid confusion in this work, ``rational'' will always refer to maximizing expected utility.

Although a rational agent always plays optimally with respect to their own expectations, the actual consequences of their actions depend on the choices of other players as well. Consequently, it is sometimes possible for a third party to regard one set of actions for all the players of a game to be superior to another set for the affected agents. As learning from experience is built into the Bayesian model, we can investigate whether such superior performance can arise from individuals single-mindedly pursuing their perceived best interests in an iterated setting. After a turn, agents can revise their priors with Bayes rule, producing a posterior distribution reflecting the outcome just obtained. If the game is played again, they can use this posterior as their new prior and potentially fare better. 

Repeated games with Bayes rule updating have been studied classically, for example to articulate circumstances when players will eventually play a Nash equilibrium~\cite{Kalai1993,Norman2022}. One immediately wonders about long term iterated play in the presence of quantum resources. Do agents typically learn to play optimally or close to optimally from the perspective of a third party? Do they learn to leverage quantum resources and outperform the best possible classical actions? If so, the consequences for algorithm design and quantum machine learning could be dramatic; if initially ignorant rational agents can autonomously discover superior play, they might be able to locate quantum algorithms which elude human intuition as well.

Explicitly considering players' beliefs not only widens the scope of discourse surrounding the games, but may produce interesting new kinds of performance phenomena reflecting the interplay between players' beliefs about a game's structure and their beliefs about one another. This setting inspires new questions. For example, can belief in a quantum resource, even in its absence, ever suffice to improve performance? Does a more sophisticated conception of a player always increase the complexity of our considerations?

In this paper, we provide Bayesian adaptations of two simple and illustrative two-player games where the addition of quantum resources is known to admit more favorable outcomes for the players: the CHSH game and the prisoners' dilemma. In \S\ref{sec:CHSH}, we present the CHSH game and then describe and simulate a version where each agent has prior beliefs about the amount of shared entanglement and the action their partner is likely to take. 
In \S\ref{sec:prisoner}, we first tell the analogous story for the prisoners' dilemma and then describe and simulate a version where the players are not only rational themselves, but also believe their opponent is rational as well. 
We conclude in \S\ref{sec:discussion} by summarizing our results and discussing future prospects for this paradigm.

\section{CHSH game}\label{sec:CHSH}
In this section, we first review the CHSH game and present the optimal classical and quantum strategies. Then we consider the same scenario from the perspective of Bayesian rational players with uncertain beliefs about the degree of entanglement between their qubit and the others' and about which action the other will take. Finally, we describe and run simulated iterations of this situation subject to four initial prior and entanglement preparation scenarios and analyze performance in terms of winning probability and entanglement expectation.

\subsection{Rules and quantum extension} The CHSH game is a version of the Clauser--Horne--Shimony--Holt experiment to test Bell inequality violations in quantum theory~\cite{Clauser1969}. Two separated players, Alice and Bob, receive uniform random bits $x$ and $y$ from a referee. Alice sends back a bit $a$ and Bob sends back a bit $b$. Then the referee awards Alice and Bob a win if
\begin{equation}\label{xor}
    x \land y = a\oplus b \;.
\end{equation}
Given they cannot communicate once the game has begun, how should they play to maximize their chances of winning?

They can win with probability 3/4 by simply always playing $a=b=0$ or $a=b=1$. Then $a\oplus b=0$ and \eqref{xor} is satisfied when $x$ and $y$ are not both $1$, which will be the case 3 times out of 4. In fact, this deterministic strategy does better than any probabilistic one based on an initially shared random variable~\cite{Wilde2017}. For this reason, 3/4 is the optimal \emph{classical} winning probability. 

The situation is different if we extend the game to allow Alice and Bob to condition their bit values on their choice of a projective measurement that the referee will make on their half of an entangled pair of particles. Specifically, we allow the referee to prepare two-particle pure states of the form 
\begin{equation}\label{possiblestates}
\ket{\psi(\gamma)}=\cos(\gamma/2)\ket{00}+\sin(\gamma/2)\ket{11}\;
\end{equation}
for any $\gamma\in[0,\pi/2]$, so that $\ket{\psi(0)}$ is separable, $\ket{\psi(\pi/2)}$ is maximally entangled, and other $\ket{\psi(\gamma)}$ have between $0$ and $1$ ebits of entanglement \cite{Bennett1996}, and allow each player to choose a measurement given by an observable 
\begin{equation}\label{possibleactions}
    \mathcal{O}(\theta,\phi)=\sin(\theta)\cos(\phi)X+\sin(\theta)\sin(\phi)Y+\cos(\theta)Z\;
\end{equation}
for any $\theta\in[0,\pi]$ and $\phi\in[0,2\pi]$, where $X$, $Y$, and $Z$ are the Pauli matrices and, for every setting, the outcomes $+1$ and $-1$ are associated with returning the bit values $0$ and $1$, respectively. 

Notice that this changes the game not only because of the presence of entanglement in the state \eqref{possiblestates}, but also due to the fact that their choice is now amongst possible quantum measurements rather than between the bit values 0 and 1 to return to the referee. This arrangement may nonetheless be thought of as an extension of the classical game because, in the separable case, each player can choose a measurement from \eqref{possibleactions} to produce the bit value $0$ or $1$ with whatever probability they wish. Furthermore, the equivalent of the optimal classical strategy is available for any amount of shared entanglement: if they both always choose $+Z$ or both always choose $-Z$, this results in equal bit values obtained by the referee and therefore they win with probability $3/4$.

As an example, suppose the referee prepares the state $\ket{\psi(\pi/3)}=\frac{\sqrt{3}}{2}\ket{00}+\frac{1}{2}\ket{11}$, where the first register is assigned to Alice and the second to Bob, and consider the following measurement strategy: If $x=0$, Alice chooses $X$, and if $x=1$, she chooses $Y$. If $y=0$, Bob chooses $X$, and if $y=1$, he chooses $(X+Y)/\sqrt{2}$. This arrangement results in a winning probability of $1/2+(\sqrt{3}+\sqrt{6})/16\approx 0.761$, just above the optimal classical winning probability.

How much can the winning probability be boosted above the classical optimum? If Alice and Bob each receive one qubit of a \emph{maximally} entangled pair, they can make their choice based on the outcome of specifically chosen projective measurements to win with probability $1/2+\sqrt{2}/4\approx 0.854$, approximately $10\%$ better than the best classical strategy. One realization of this rate is with the state $\ket{\psi(\pi/2)}=\frac{1}{\sqrt{2}}(\ket{00}+\ket{11})$ and the following measurement strategy: If $x=0$, Alice chooses $Z$, and if $x=1$, she chooses $X$. If $y=0$, Bob chooses $(X+Z)/\sqrt{2}$, and if $y=1$, he chooses $(Z-X)/\sqrt{2}$. 

As it is particularly simple and its quantum extension straightforwardly improves the players' situation, the CHSH game furnishes a convenient setting for us to look at through a Bayesian lens in the coming subsections. 

\subsection{Rational players}\label{sec:rationalplayers}
What does it take to actually achieve optimal play in this game? One way would be for Alice and Bob to meet ahead of time to decide their joint strategy, for example, always choosing $a=b=0$ in the classical version. Another might be for Alice to announce her intended strategy and then to rely on the heuristic reasoning capabilities of Bob to play in the appropriate complementary way. And what happens if the state the referee assigns to the players' systems is unknown to the players? In a general game scenario, each player faces a decision which they can only make on the basis of their uncertain knowledge and beliefs. As indicated in the Introduction, we seek to understand how Bayesian rational agents fare in the CHSH game.

We adopt the convention that when more than one action maximizes a rational agent's expected utility, they tiebreak by choosing uniformly at random among the maximizing options. We also do not concern ourselves with the origin of any particular prior. An actual agent might arrive at initial beliefs in any number of ways, for example, by considering their past experiences or internal models for the behavior of devices and the actions and reasoning of other agents, but the formal apparatus of decision theory does not impose any particular systematic prescription.  As the CHSH game only has two outcomes, winning or losing, we consider their expected utility equivalent to their expected winning probability.

First consider the original, non-extended CHSH game, but now Alice and Bob are Bayesian rational agents. Suppose Alice is sure Bob will play 0 regardless of the bit he receives and Bob is sure Alice will play 1 regardless of the bit she receives. These priors, while perhaps uncharitable to the reasoning of the other, are possible for rational agents to hold because we have not assumed that either necessarily believes in the rationality of the other or ascribes any more nuanced attributes to the others' reasoning. If Alice receives $x=0$, the left hand side of \eqref{xor} is zero and so they win only if $a$ and $b$ are the same. Since she is sure Bob will play $b=0$, she chooses $a=0$. If she instead receives $x=1$, she expects to win half the time for either $a$ value. The situation is similar for Bob. Combining their rational actions for each of the 4 possible values of $x$ and $y$ reveals that their priors together amount to a rational strategy which wins with probability $3/8$, not only inferior to the optimal rate, but also worse than the winning probability of $1/2$ achieved by choosing $a$ and $b$ uniformly at random. 

These agents are rational, but, due to their strong beliefs, lose more often than they win. In fact, as they are \emph{certain}, they fail to learn anything from Bayes rule updating. This is a simple consequence of the fact that Bayes rule produces posteriors proportional to the prior and the likelihood, where a likelihood is the probability the agent assigns to a particular outcome conditioned on the truth of one of the hypotheses of the prior. If any hypothesis has zero prior weight, it will have zero posterior weight as well. Likewise, if an agent considers an outcome to be impossible given a particular hypothesis, then, if that outcome does occur, their subsequent posteriors will have zero weight for that hypothesis. 

As it excludes learning in this way, some Bayesians have argued that one should \emph{never} assign a prior probability of zero or one to anything other than purely logical propositions, elevating this dictum to a norm of rationality that Lindley called Cromwell's rule~\cite{Lindley1985}. We argue, however, that completely forbidding certainty unduly strains one's decision making. If no probability should ever be zero, one's sample space should always be infinite, regardless of what they believe to be relevant. Moreover, the coherence of Bayes updating itself actually rests on an instance of certainty. Specifically, Jeffrey's probability kinematics only reduces to Bayesian conditioning if one believes that their future beliefs will be determined with certainty by the appearance of some relevant data (see section 5 of \cite{Fuchs2012}).\footnote{We thank Christopher Fuchs and Blake Stacey for these observations.} 

As we will soon see, rational players who are at least a little uncertain stand a much better chance of learning from experience when the game is repeated many times. 

In the extended CHSH game, both players know that they will receive a bit value uniformly at random and that they get to choose an observable \eqref{possibleactions} that the referee will measure for one of a pair of potentially entangled particles to determine the bit values $a$ and $b$. Furthermore, given their own bit value, the two qubit game state prepared by the referee, and their partner's measurement strategy, each knows how to use the Born rule and the winning condition \eqref{xor} to calculate the winning probability of any of their own possible actions. They do not know, however, which action the other will actually take or which game state the referee will prepare. This amounts to uncertainty about, and therefore prior probabilities for, the parameters $\theta$ and $\phi$ chosen by their partner and the entanglement parameter $\gamma$ in the game state. 

Winning or losing is actually determined by the outcomes of the referee's measurements, so when we talk about the  winning probability, we mean the probability that the referee assigns to satisfying the winning condition given their state and the prescribed subsystem measurements from Alice and Bob. Alice and Bob, of course, do not have access to the referee's winning probability; they reason using their current priors for the action of the other and the amount of entanglement in the game state and act such that their personal expected winning probability is maximized.

The extended game is sufficiently complex that we resort to numerical simulation to explore the effects of agents' beliefs on game outcomes. Our simulations and their results are described in the next subsection.

\subsection{Simulated iterations}

In this subsection we describe and discuss some simple simulations of repeated rounds of the extended CHSH game starting from particular priors and game state preparations and where the state of knowledge of each player changes according to Bayes rule between each round. 

Each round that we simulate consists of \emph{three} iterations of the game, that is, the agents play the game three times per round with rational actions chosen from their current prior. These repetitions, while not strictly necessary, generally expose the players to different initial bit values, allowing them to learn from the consequences of rational actions reflecting a broader portion of their current beliefs. After each round, they are told the bit values that the other agent received for each iteration, but they are \emph{not} told which actions their opponent took. Each then updates their prior to a posterior with Bayes rule in light of the pair of bit values, the action they personally took, and the outcome, win or lose, for each of the three iterations. It is important to realize that we are assuming there are no memory effects, as in \cite{Barrett2002}, and that each round consists of three separate wins or losses, rather than a win only if they win all three and a loss otherwise, as in \cite{Cleve2008}. Agents' beliefs pertain to single rounds and between rounds they take nothing with them but their updated prior. Consequently, they do not know how many rounds have elapsed nor how many rounds are yet to be played.

\subsubsection{Discretizations and probability floor}
\label{chshdiscretizations}
To simulate rounds of the CHSH game we fix two further details. First, to make simulations computationally tractable, we impose discretizations of the spaces of possible shared quantum states \eqref{possiblestates} and actions \eqref{possibleactions}. We consider 11 $\gamma$ values associated to entanglement ranging from $0$ to $1$ ebits\footnote{Although the parameter $\gamma$ in the game state definition is unitless, we refer to it in our simulations in terms of the number of ebits in the associated state.} in steps of $0.1$. The ranges for $\theta$ and $\phi$ are divided into steps of $\pi/8$: 9 values for $\theta$ from $0$ to $\pi$ and 16 values for $\phi$ from $0$ to $15\pi/8$. When $\theta$ equals $0$ or $\pi$, the $\phi$ value does not change the measurement, so we remove this redundancy by only including one of them. 

Second, we insert a constant nonzero probability floor for all possible measurements. To do this, we make the substitution $P\rightarrow (1-\epsilon)P+\frac{\epsilon}{2}I$ for each measurement projector $P$ associated with the observable $\mathcal{O}(\theta,\phi)$ and choose $\epsilon=0.1$.\footnote{We found this $\epsilon$ value allowed for expressive simulations within a manageable number of rounds.} The way we have chosen $\epsilon$ is arbitrary, but some regularization of this kind is important to avoid the pitfall of certainty mentioned above --- for perfectly pure states and perfect projective measurements, some outcomes will be given precisely zero probability for certain hypotheses. If such an outcome occurs, the agent would thereafter regard certain hypotheses to be impossible and strongly limit their learning potential. The scheme we have chosen corresponds to the situation of a measurement with depolarizing noise.

The probability floor globally changes winning probabilities in the game, but this is acceptable because only relative values are important for us and our constant floor does not affect this. For $\epsilon=0.1$, the optimal winning probability without entanglement is now modified to $1/2+(1-\epsilon)^2/4\approx 0.703$ instead of $0.750$ and the optimal winning probability with maximal entanglement is $1/2+\sqrt{2}(1-\epsilon)^2/4 \approx 0.786$ instead of $\approx 0.854$. 

\subsubsection{An example round}

Let's walk through one round of this game. Suppose the game state is maximally entangled and that both Alice and Bob have a completely uniform initial prior over the discretized action and entanglement sample space for both bit values of the other. Recall that we regard a single round to be three iterations of the game, which means that each player receives three bit values. Suppose Alice receives the bits $x=\{0,0,1\}$ for the three iterations of this round and Bob receives $y=\{0,1,1\}$. As they regard every action of the other to be equally likely, irrespective of the bit value the other receives, they expect every action to win with probability 0.5. Consequently, they randomly pick a measurement to take each time they receive a $0$ and randomly pick one to take each time they receive a $1$. Alice chooses $\mathcal{O}(7\pi/8,7\pi/4)$ for $x=0$ and $\mathcal{O}(\pi/4,5\pi/4)$ for $x=1$ while Bob chooses $\mathcal{O}(\pi/8,15\pi/8)$ for $y=0$ and $\mathcal{O}(\pi/4,\pi/4)$ for $y=1$. Given the uniform probabilities over the initial bit values, the joint winning probability of this strategy is 0.371. For the actual bit values received during the three iterations, a win is simulated with probabilities $\{0.177,0.345,0.298\}$. Suppose the most likely outcomes occur and they lose all three iterations. On the basis of the bit values, outcomes, and the actions they took, each updates their prior with Bayes rule where their likelihood is constructed in the natural way from the rules of game.

After the first round, Alice's beliefs about Bob's actions have shifted towards smaller $\theta$ values and mid-range $\phi$ values for either bit value he could receive; the largest probabilities she assigns are for the actions $\mathcal{O}(\pi/8,5\pi/4)$ if $y=0$ and $\mathcal{O}(\pi/8,7\pi/8)$ if $y=1$. Meanwhile Bob's beliefs about Alice's actions have shifted to higher $\theta$ values with mid-range $\phi$ values if she receives $x=0$, but to lower $\theta$ values and higher $\phi$ values if she receives $x=1$. The measurements he considers her most likely to take are $\mathcal{O}(7\pi/8,7\pi/8)$ if $x=0$ and $\mathcal{O}(\pi/8,7\pi/4)$ if $x=1$. For both of them, these shifts are relatively small --- they initially assigned probability $\approx 0.009$ to every action, while after updating, the largest probability either assigns to any action is $\approx 0.021$. After this single round, their entanglement marginal posterior predictions change imperceptibly. 

Going into the second round, Alice and Bob now expect to win with probability $0.600$ and $0.598$, respectively, but, in fact, the joint winning probability entailed by their new beliefs and the game state actually lowers their winning chances to $0.298$. This certainly does not seem to be a step in the right direction. Is rationality simply a losing strategy? After many rounds, we will see that it is not.

\subsubsection{Simulation scenarios}\label{sec:chshSIMs}

\begin{figure*}
     \centering
     \begin{subfigure}[t]{0.49\textwidth}
         \centering
         \includegraphics[width=\textwidth]{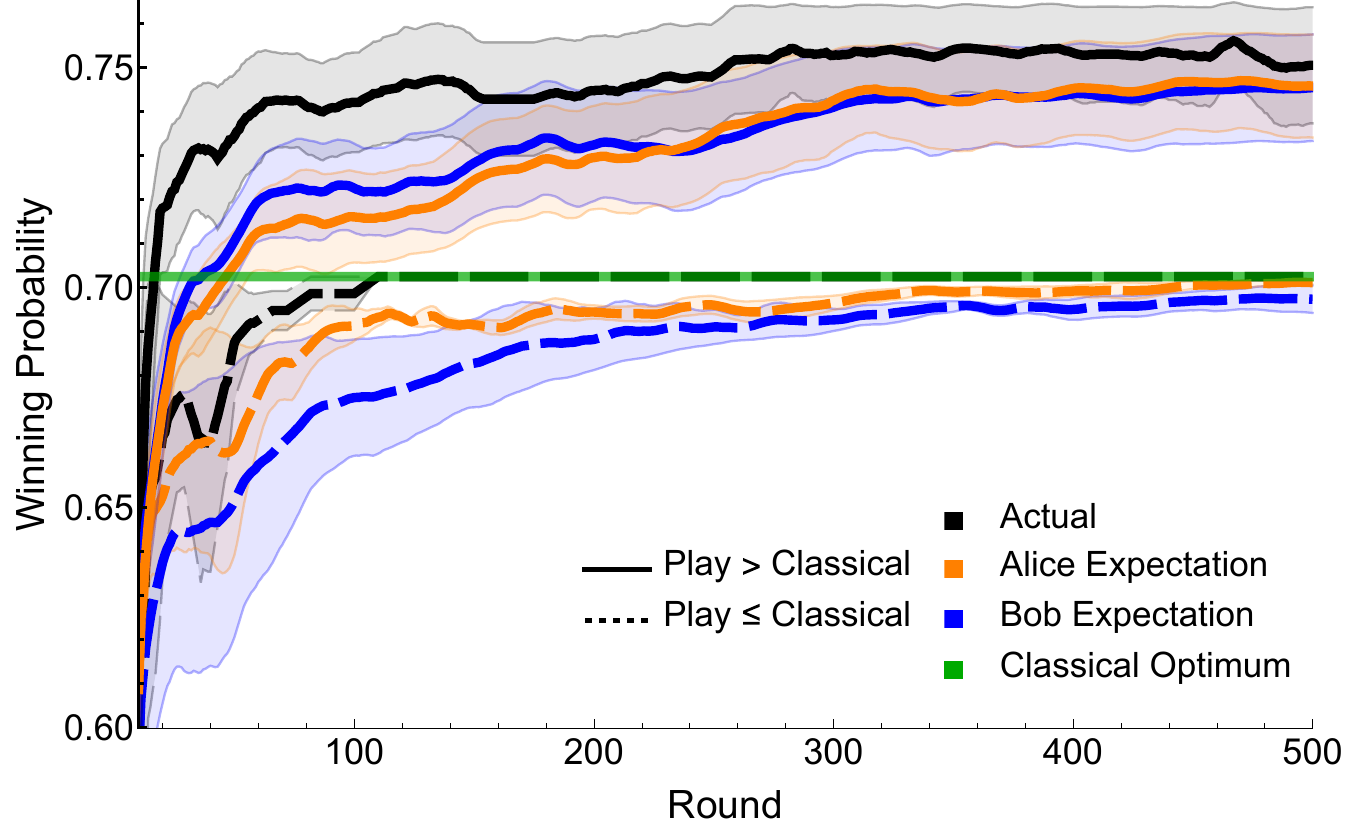}
         \caption{CHSH Scenario \ref{chsh1} (Finding Advantage): $\gamma_g=1.0$ ebits, Alice (orange) and Bob (blue) both start with the \hyperlink{uniform}{\color{black}uniform} prior. Performance (black) rises in all simulations to one of two relatively stable outcomes. In 6 simulations, averaged as solid lines with standard deviation bands, play exceeds the classical optimum and reaches an average winning probability above 0.75. In the other 4, averaged as dashed lines with standard deviation bands, winning probability initially rises more slowly and stabilizes precisely at the classical optimum. In all cases, Alice and Bob's expectations lag behind the actual winning probability, but have mostly closed the gap by the final rounds. As the players' situation is symmetric in this scenario, statistical discrepancies between them are due to running only 10 simulations.}
         \label{subfig:chsh1}
     \end{subfigure}
     \hfill
     \begin{subfigure}[t]{0.49\textwidth}
         \centering
         \includegraphics[width=\textwidth]{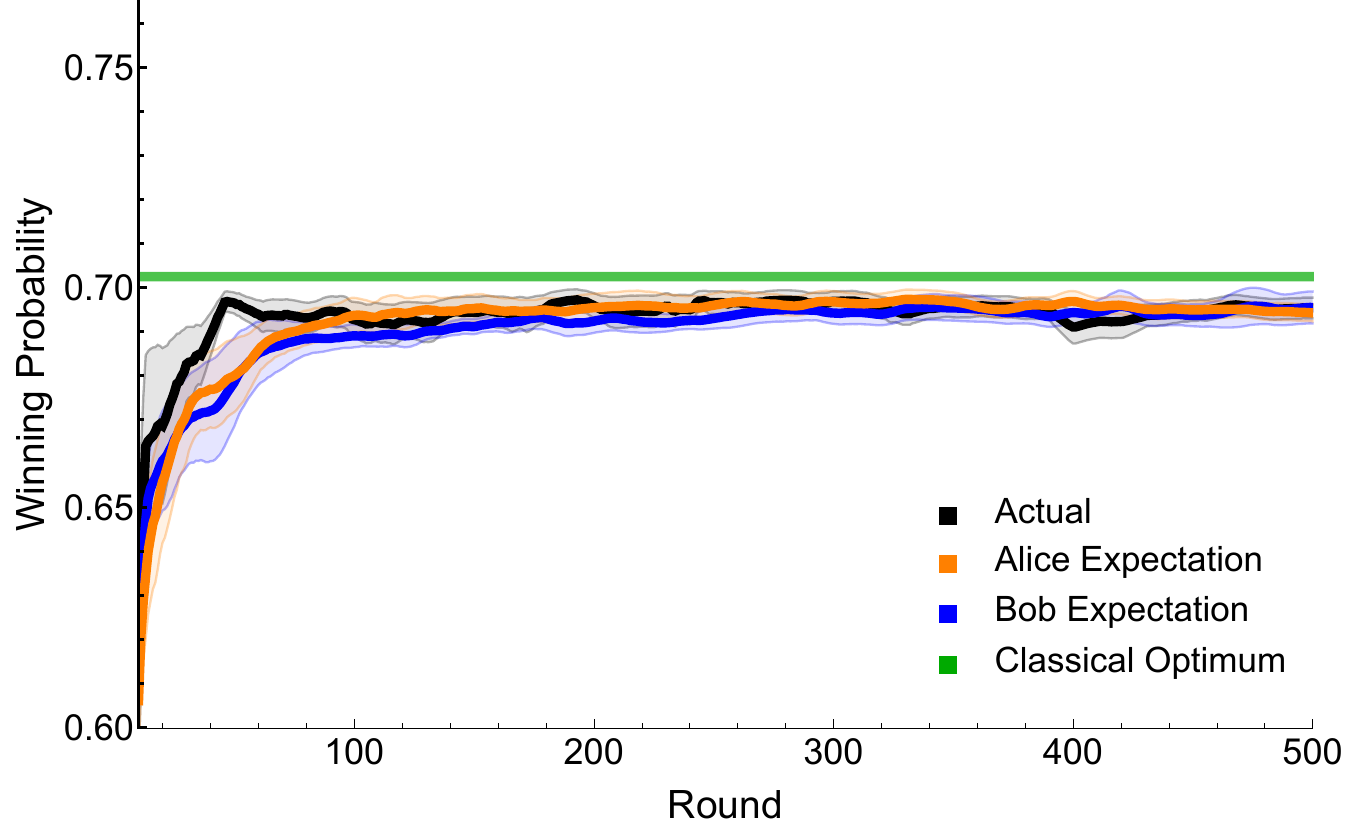}
         \caption{CHSH Scenario \ref{chsh2} (Making Do): $\gamma_g=0.0$ ebits, Alice (orange) and Bob (blue) both start with the \hyperlink{uniform}{\color{black}uniform} prior. Performance (black) and expected performance rise in step with each other in all simulations until converging to or near the classical optimum. Standard deviation bands are plotted around each average. In 6 simulations they play precisely the classical optimum in a plurality of rounds, while in the other 4 they play with just under the classical optimum in the majority of rounds. After an average of 150 rounds, the average actual winning probability as well as Alice and Bob's expectations arrive and remain within 0.01 of the classical optimum. As the players' situation is symmetric in this scenario, statistical discrepancies between them are due to running only 10 simulations.}
         \label{subfig:chsh2}
     \end{subfigure}
     \hfill
     \begin{subfigure}[t]{0.49\textwidth}
         \centering
         \includegraphics[width=\textwidth]{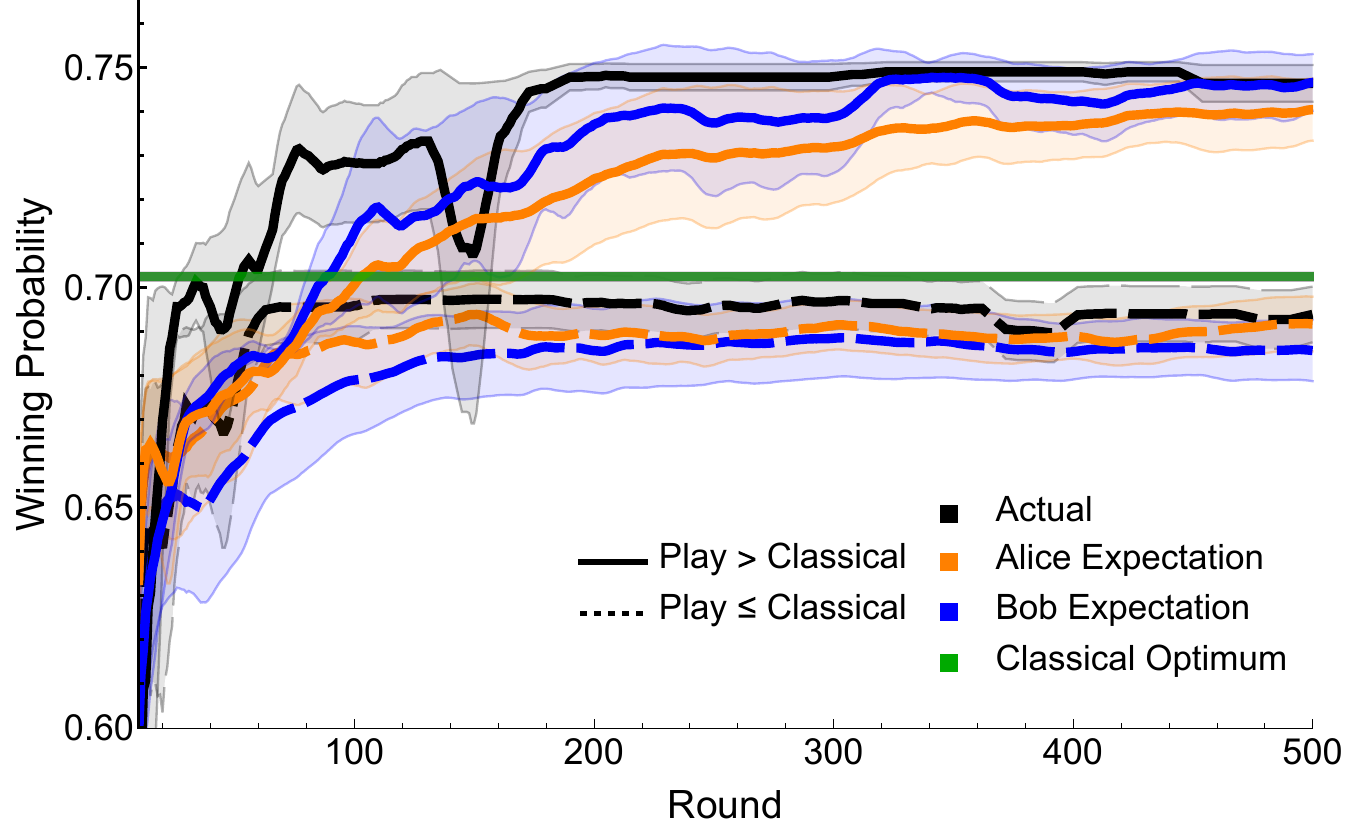}
         \caption{CHSH Scenario \ref{chsh3} (Overcoming Bias): $\gamma_g=0.7$ ebits, Alice (orange) starts with the \hyperlink{skewC}{\color{black}skew classical} prior and Bob (blue) starts with the \hyperlink{uniform}{\color{black}uniform} prior. Performance (black) rises in all simulations to one of two stable outcomes. In 3 simulations, averaged as solid lines with standard deviation bands, play exceeds the classical optimum and reaches an average winning probability just under 0.75. In the other 7, averaged as dashed lines with standard deviation bands, winning probability initially rises more slowly and stabilizes just below the classical optimum on average. While both players' expectations track well with the actual performance across all simulations, Bob's tracks closer to actual when play exceeds the classical optimum and Alice's tracks closer when it does not.}
         \label{subfig:chsh3}
     \end{subfigure}
     \hfill
     \begin{subfigure}[t]{0.49\textwidth}
         \centering
         \includegraphics[width=\textwidth]{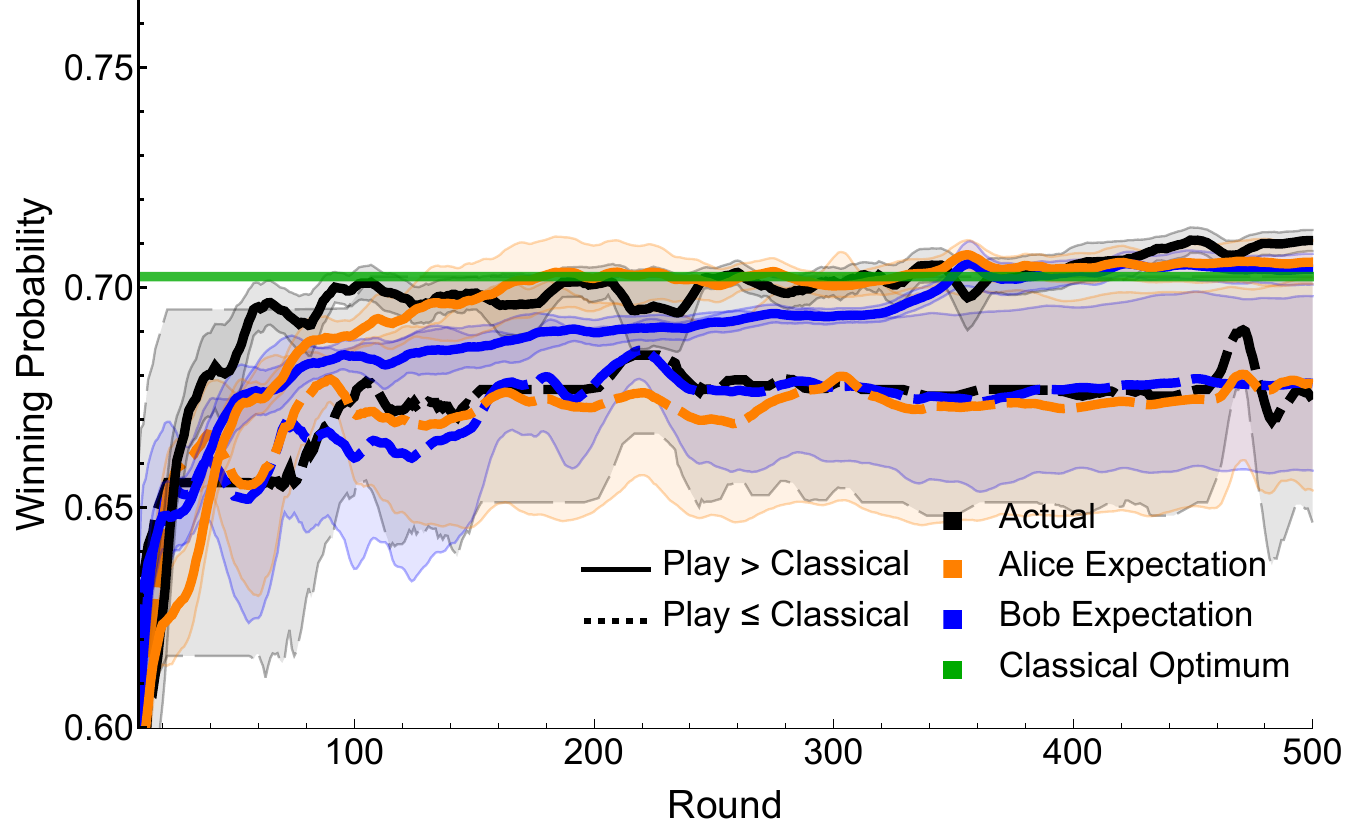}
         \caption{CHSH Scenario \ref{chsh4} (Good Enough?): $\gamma_g=0.3$ ebits, Alice (orange) starts with the \hyperlink{skewQ}{\color{black}skew quantum} prior and Bob (blue) starts with the \hyperlink{uniform}{\color{black}uniform} prior. Performance (black) rises in all simulations and exhibits one of two trends. In 7 simulations, averaged as solid lines with standard deviation bands, after an initial rise, winning probability creeps higher, only exceeding the classical optimum on average in the final 100 rounds. In the other 3, averaged as dashed lines with standard deviation bands, winning probability rises slowly and stabilizes significantly below the classical optimum at an average winning probability under 0.68. When they manage to exceed the classical optimum, Alice's prior leads her to predict a quantum advantage almost 200 rounds before Bob's does.}
         \label{subfig:chsh4}
     \end{subfigure}
        \caption{Winning probability (black) and players' expected winning probabilities for simulations of rational agents Alice (orange) and Bob (blue) playing 10 simulations of 500 3-iteration rounds of the CHSH game for each of the four prior and game state entanglement scenarios described and analyzed in \S\ref{sec:chshSIMs}. The classical optimum (green) is $0.703$ rather than $0.750$ due to the probability floor discussed in \ref{chshdiscretizations}. In each subfigure, we plot the 10-round moving average and display the standard deviation of the mean with uncertainty bands. After each round, both players update their single-round priors with Bayes rule before the next round begins. }
        \label{fig:CHSHperformance}
\end{figure*}

\begin{figure*}
     \centering
     \begin{subfigure}[t]{0.49\textwidth}
         \centering
         \includegraphics[width=\textwidth]{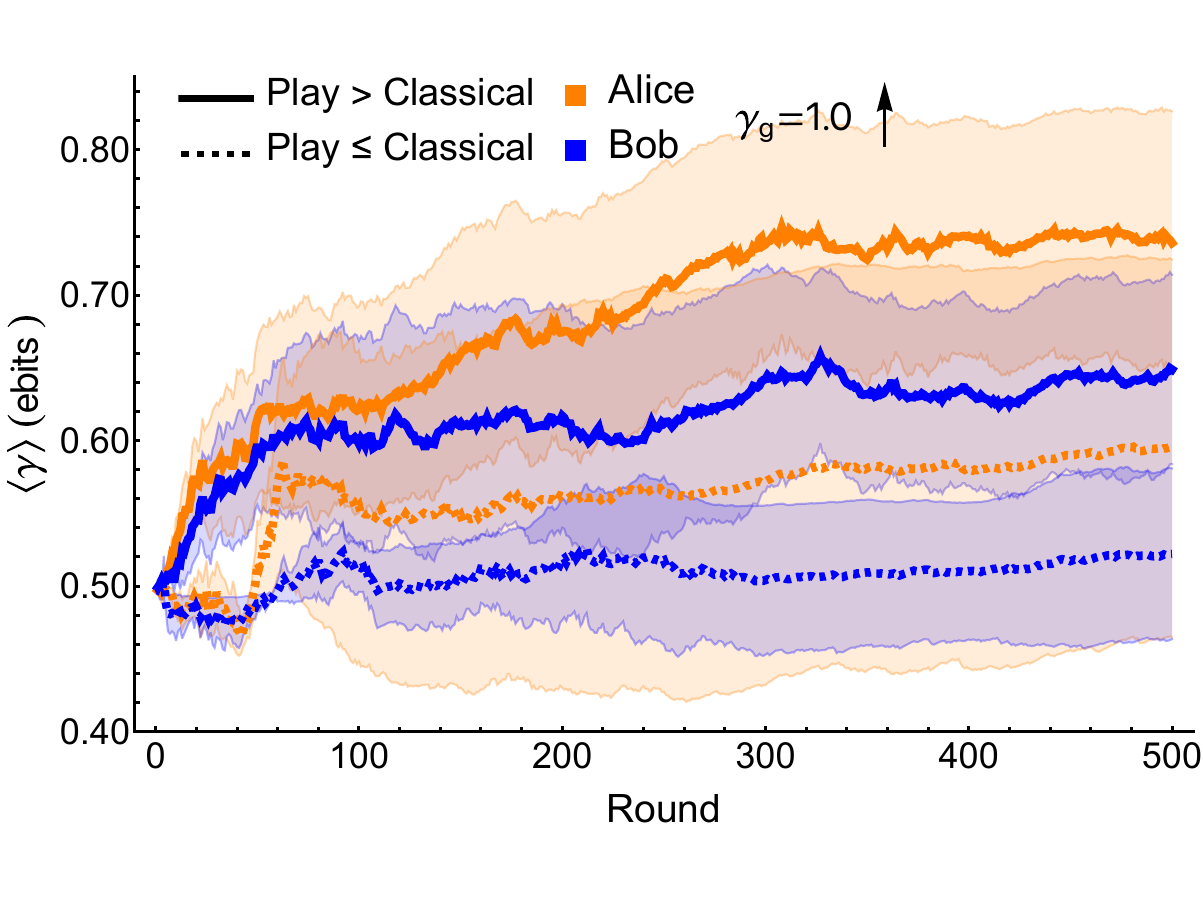}
         \caption{CHSH Scenario \ref{chsh1} (Finding Advantage): $\gamma_g=1.0$ ebits (above plotted range), Alice (orange) and Bob (blue) both start with the \hyperlink{uniform}{\color{black}uniform} prior. The solid lines are the average with standard deviation bands for the 6 simulations where play exceeded the classical optimum and the dotted lines are the average and standard deviation for the remaining 4 simulations. After the initial rise in winning probability, players' entanglement expectations notably rise on average when play manages to exceed the classical optimum and much less on average when it does not. However, variation in these trends is high. As the players' situation is symmetric in this scenario, statistical discrepancies between them are due to running only 10 simulations. }
         \label{subfig:chshENT1}
     \end{subfigure}
     \hfill
     \begin{subfigure}[t]{0.49\textwidth}
         \centering
         \includegraphics[width=\textwidth]{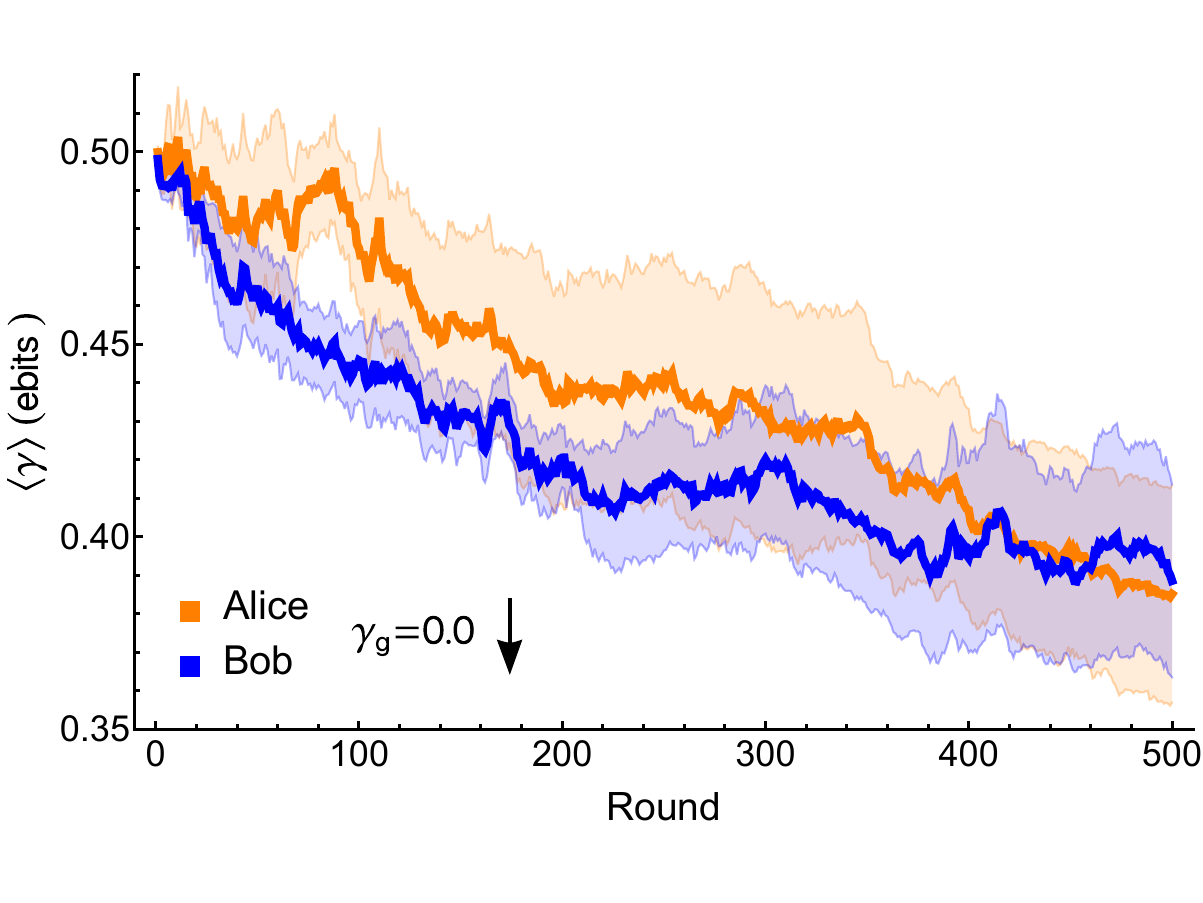}
         \caption{CHSH Scenario \ref{chsh2} (Making Do): $\gamma_g=0.0$ ebits (below plotted range), Alice (orange) and Bob (blue) both start with the \hyperlink{uniform}{\color{black}uniform} prior. Lines indicate the average with standard deviation bands for all simulations. Entanglement expectation falls with the rise in winning probability, but has only modestly reduced on average by the final rounds. A reason for this is that the same classical optimal winning value may be obtained by many combinations of measurements and entanglements. As the players' situation is symmetric in this scenario, statistical discrepancies between them are due to running only 10 simulations.}
        \label{subfig:chshENT2}
     \end{subfigure}
     \hfill
     \begin{subfigure}[t]{0.49\textwidth}
         \centering
         \includegraphics[width=\textwidth]{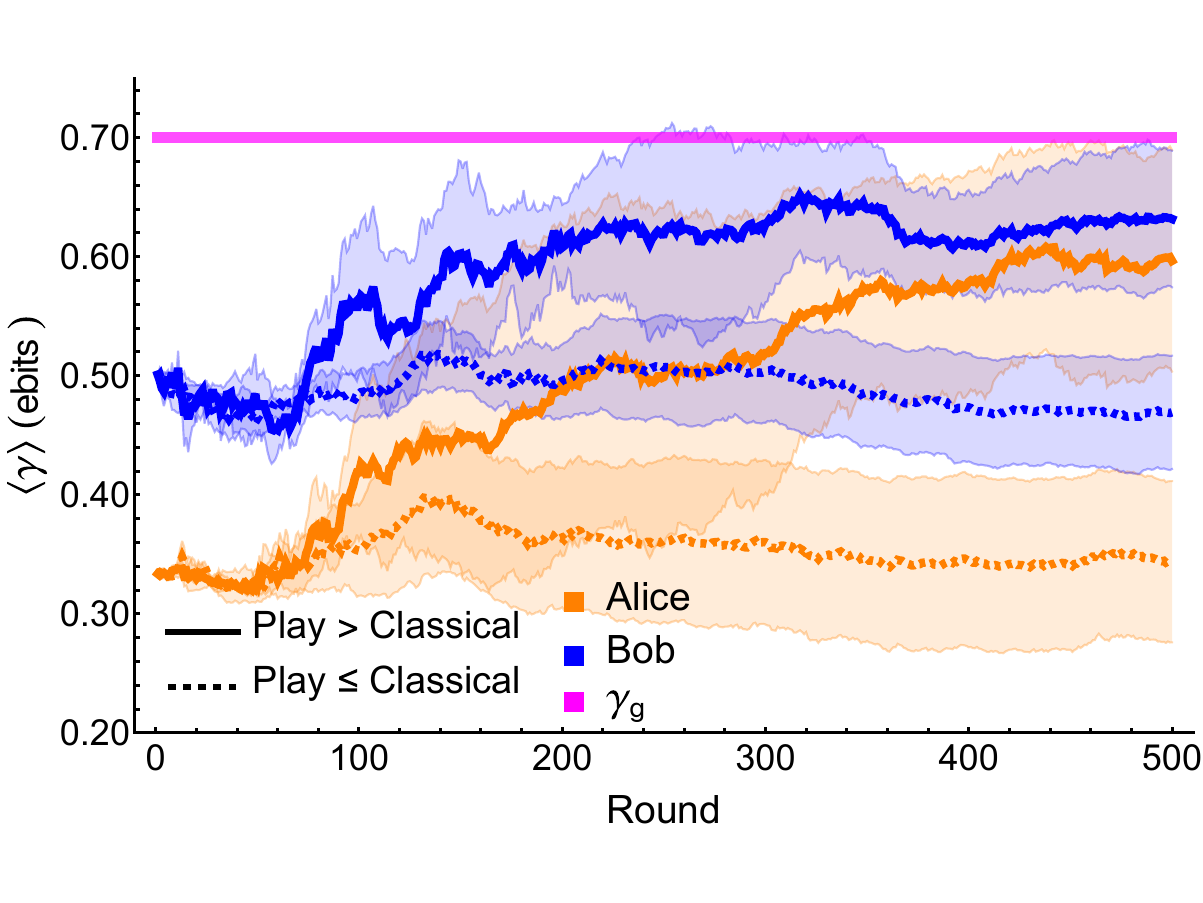}
         \caption{CHSH Scenario \ref{chsh3} (Overcoming Bias): $\gamma_g=0.7$ ebits (pink), Alice (orange) starts with the \hyperlink{skewC}{\color{black}skew classical} prior and Bob (blue) starts with the \hyperlink{uniform}{\color{black}uniform} prior. The solid lines are the average with standard deviation bands for the 3 simulations where play exceeded the classical optimum and the dotted lines are the average and standard deviation for the remaining 7 simulations. After the initial rise in winning probability, players' entanglement expectations notably rise on average when play manages to exceed the classical optimum, but exhibits no such trend when it does not. In the first situation, Alice's expectations catch up with Bob's on average by the final rounds. }
         \label{subfig:chshENT3}
     \end{subfigure}
     \hfill
     \begin{subfigure}[t]{0.49\textwidth}
         \centering
         \includegraphics[width=\textwidth]{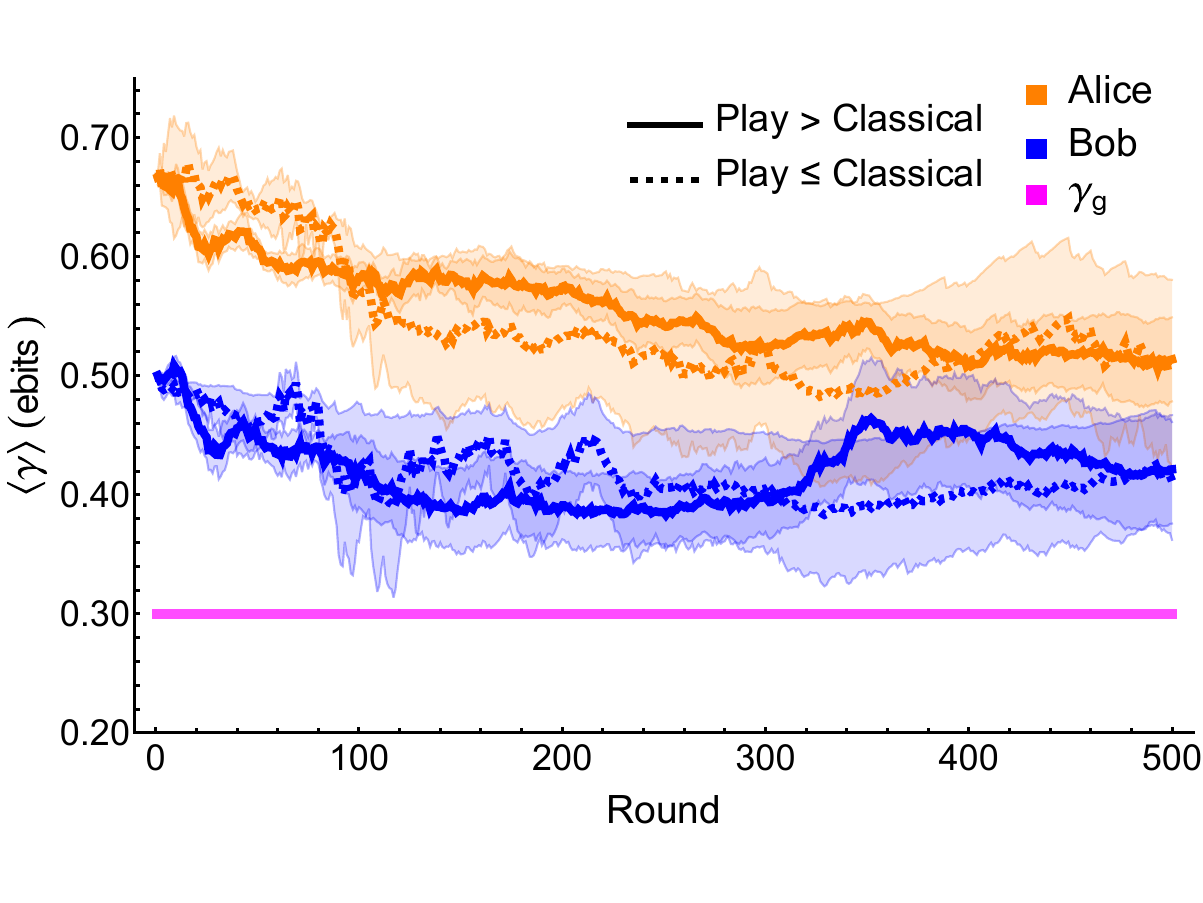}
         \caption{CHSH Scenario \ref{chsh4} (Good Enough?): $\gamma_g=0.3$ ebits (pink), Alice (orange) starts with the \hyperlink{skewQ}{\color{black}skew quantum} prior and Bob (blue) starts with the \hyperlink{uniform}{\color{black}uniform} prior. The solid lines are the average with standard deviation bands for the 7 simulations where play exceeded the classical optimum and the dotted lines are the average and standard deviation for the remaining 3 simulations. Entanglement expectation falls with the rise in winning probability in all simulations. In the number of rounds simulated, the simulations where play eventually manages to exceed the classical optimum and those where it does not exhibit indistinguishable entanglement expectation trends.}
         \label{subfig:chshENT4}
     \end{subfigure}
        \caption{Entanglement expectations for simulations of rational agents Alice and Bob playing 10 simulations of 500 3-iteration rounds of the CHSH game for each of the four prior and game state entanglement scenarios described and analyzed in \S\ref{sec:chshSIMs}. Bands around values represent one standard deviation. After each round, both players update their single-round priors with Bayes rule before the next round begins. }
        \label{fig:chsh_entanglement}
\end{figure*}

We consider four scenarios of this game which we call ``Finding Advantage'', ``Making Do'', ``Overcoming Bias'', and ``Good Enough?''. For each, we ran 10 simulations of 500 3-iteration rounds. We summarize our results in Figures \ref{fig:CHSHperformance} and \ref{fig:chsh_entanglement}. Figure \ref{fig:CHSHperformance} tracks winning probability and players' expected winning probability by round and Figure \ref{fig:chsh_entanglement} tracks how players' entanglement expectations evolve. 

In the scenarios below we fix a game state $\gamma_g$ value\footnote{We speak of \emph{the} game state or of the state the referee prepares for brevity, but the reader should remember that it is the referee's personal quantum state, in accordance with the QBist framework we have adopted.} and we initialize each agent with one of three priors:

\begin{uniform}\hypertarget{uniform}The agent has a uniform prior over the full action space of the other for each bit value the other could receive. The agent's entanglement prior is also uniform.
\end{uniform}

\begin{skewC}\hypertarget{skewC}For both bit values the other could receive, the entanglement prior is the sum-normalized list of consecutive integers descending from 11 to 1, such that the highest probability is $1/6$ for 0 ebits and the initial expected entanglement is $1/3$ ebits. As a formula, $P(n\times 0.1\text{ ebits})=(11-n)/66,\; n\in\{0,\ldots,10\}$. The prior over the set of measurements has two peaks at $+Z$ and $-Z$, that is, $\theta=0$ and $\theta=\pi$, and uniform over the allowed $\phi$ values. Explicitly, the $\theta$ marginal is proportional to the nine-element list $\{e^5,e^4,e^3,e^2,e,e^2,e^3,e^4,e^5\}$. We say this prior is ``classically'' skewed because it is biased towards lower entanglement and the classical optimum actions of $+Z$ and $-Z$. 
\end{skewC}

\begin{skewQ}\hypertarget{skewQ}For both bit values the other could receive, the entanglement prior is the sum-normalized list of consecutive integers ascending from 1 to 11, such that the highest probability is $1/6$ for 1 ebits and the initial expected entanglement is $2/3$ ebits. As a formula, $P(n\times 0.1\text{ ebits})=n/66,\; n\in\{0,\ldots,10\}$. The prior over measurements is peaked at $\theta=\pi/2$ and uniform over the allowed $\phi$ values. Explicitly, the $\theta$ marginal is proportional to the nine-element list $\{e,e^2,e^3,e^4,e^5,e^4,e^3,e^2,e\}$. We call this prior ``quantum'' skewed because it is biased towards higher entanglement and away from the ``classical'' measurement actions of $+Z$ and $-Z$. 
\end{skewQ}

\begin{chsh}[Finding Advantage]\label{chsh1}
This is the scenario that we walked through one round of above. The referee prepares the maximally entangled game state, $\gamma_g=1.0$ ebits, and Alice and Bob both have the \hyperlink{uniform}{\color{black}uniform} initial prior. This is perhaps the most basic scenario that the entangled state makes possible: Both players believe in some entanglement and have a blank-slate prior for the other. Will they find the quantum advantage?

As shown in Figure \ref{subfig:chsh1}, performance rises in all simulations to one of two relatively stable outcomes. In 6 simulations, winning probability exceeds the classical optimum after an average of 20 rounds and players' expectations follow another 20 rounds later. In the other 4 simulations, winning probability meets and stabilizes at precisely the classical optimal value and players' expectations converge towards it from below in the remaining rounds.  

Figure \ref{subfig:chshENT1} reveals that in the simulations where winning probability exceeds the classical optimum, the players' expectations are also more sensitive to the game state entanglement. However, variation in this trend is high as indicated by the standard deviation bands. This spread occurs because belief in entanglement is not the only factor controlling performance: It is possible for players to exceed the classical optimum without a strong belief in entanglement and also possible for them to fail to do so despite having a strong belief in entanglement.

These observations may be puzzling. For example, if Alice and Bob both expect a large amount of entanglement and the game state is maximally entangled, how could they fail to exceed the classical optimum winning probability? The answer is that by this time they may have become so confident that the other will fail to play a measurement which can produce a quantum advantage that their own expected payoff for a classically optimal measurement exceeds any which could have utilized the entanglement. This suggests that actions and consequences in early rounds can strongly affect long term learning prospects for agents without sharp initial beliefs. More generally, one can see that beliefs about other agents can negate quantum advantage.
\end{chsh}

\begin{chsh}[Making Do] \label{chsh2} The referee prepares the separable game state, $\gamma_g=0.0$ ebits, and Alice and Bob both have the \hyperlink{uniform}{\color{black}uniform} initial prior. The only difference between this scenario and the first is that now no quantum resource is available because the game state is separable. Will they make do with what they've got and find the classical optimal play?

As shown in Figure \ref{subfig:chsh2}, winning probability and players' expected winning probability rises in all simulations until converging to or near the classical optimum. Without entanglement in the game state, the winning probability is, of course, bounded above by the classical optimum value. Players' expectations are not bounded and in half of the simulations, for a few rounds, one or the other player does expect to win with a higher probability than the classical optimum before being suppressed by subsequent outcomes. This effect is not significant enough to see within one standard deviation of the mean, however.

Figure \ref{subfig:chshENT2} shows that entanglement expectation falls with the rise in winning probability, but only an average of about 0.1 ebits across all rounds. In none of the simulations does either agent end up with a sharp prediction of very low entanglement although expectation has not equilibrated by the end of the simulation at 500 rounds. Even in the most extreme cases, final posteriors have an expected 0.2 ebits of entanglement. A reason for this is that the same classical optimal winning value may be obtained by many combinations of measurements and entanglements, such that, in many cases, bit values and game outcomes are not enough to make higher entanglement values less plausible to our agents. 
\end{chsh}

\begin{chsh}[Overcoming Bias] \label{chsh3}The referee prepares the 0.7 ebit game state, $\gamma_g=0.7$ ebits, Alice has the \hyperlink{skewC}{\color{black}skew classical} initial prior, and Bob has the \hyperlink{uniform}{\color{black}uniform} initial prior. There is relatively high entanglement, but Alice thinks it's low; she's also biased towards actions for Bob which he might prefer if there were low entanglement. Can Alice overcome her biases?

We can see see from Figure \ref{subfig:chsh3} that winning probability and players' expected winning probabilities rise in all simulations to one of two relatively stable outcomes. In 3 simulations, winning probability exceeds the classical optimum after an average of about 60 rounds and reaches an average winning probability just under 0.75. Both players' expected winning probabilities also cross the classical threshold shortly after round 100, with Bob's expectations crossing first on average. Thereafter, Alice's winning probability estimate trails Bob's by more than when we considered agents with the same initial prior. Near round 150 we also see a dip in the average winning probability; this is because in one simulation, winning probability briefly fell back below the classical optimum. In the other 7 simulations, winning probability and players' expectations do not cross the classical optimum boundary. After the initial rise, winning probability stabilizes an average of about 0.01 below the classical optimum. Alice's expected winning probability settles slightly further below. 

In Figure \ref{subfig:chshENT3}, as initial priors are different, initial entanglement expectation is different for each agent. Similar to Scenario \ref{chsh1} (Finding Advantage), after the initial rise in winning probability, players' entanglement expectations notably rise on average when winning probability manages to exceed the classical optimum. Although Alice starts further from the game state, her expectations have caught up with Bob's on average by the final rounds and both are within 0.15 ebits of $\gamma_g$. When winning probability does not ever exceed the classical optimum, we see no significant signature of entanglement detection. 
\end{chsh}

\begin{chsh}[Good Enough?]\label{chsh4} The referee prepares the 0.3 ebit game state, $\gamma_g=0.3$ ebits, Alice has the \hyperlink{skewQ}{\color{black}skew quantum} initial prior, and Bob has the \hyperlink{uniform}{\color{black}uniform} initial prior. The actual game state entanglement is low, lower than both players expect. Alice actually thinks it's high. Is a situation like this good enough for our players to find a quantum advantage?

Figure \ref{subfig:chsh4} exhibits a final example of two trends across our simulations. In 7 simulations, after a quick initial rise, winning probability slowly grows on average and finally slightly exceeds the classical optimum after 400 rounds. Here, due to her initial belief in a large amount of entanglement, Alice's expected winning probability exceeds the classical optimum before she and Bob are actually winning with probabilities in this region. Bob's expected winning probability lags behind the actual as in other scenarios. In the other 3 simulations, winning probability rises more slowly and stabilizes significantly below the classical optimum at an average winning probability under 0.68. Both Alice and Bob's expectations track well with the actual in this situation.

As shown in Figure \ref{subfig:chshENT4}, unlike in Scenarios \ref{chsh1} (Finding Advantage) and \ref{chsh3} (Overcoming Bias), here we do not see signatures of whether winning probability exceeds the classical optimum in the players' entanglement expectations in the rounds simulated. In other words, whether agents manage to surpass the classical optimum or not does not seem to be strongly connected to whether their entanglement expectations tracked well with the game state entanglement although it does appear that the players are stably learning that the entanglement is low regardless of their performance. Because the players' entanglement expectations exceeded the actual entanglement, it may be that the degree of difference is not imparted. 

\end{chsh}

Our simulation scenarios allow us to begin addressing our questions from the introduction. Indeed, at least for the CHSH game, initially ignorant Bayesian agents \emph{can} learn to surpass classically optimal play through repeated revisions of their beliefs over the course of many iterations of a quantum game. When superior play was achieved, we usually found that players' expectations gradually shifted on average to reflect a stronger belief in the presence of entanglement, the relevant quantum resource for this game. 

We also identify some interesting features that can arise when explicitly modeling players' beliefs. In particular, the inclusion of beliefs about the others' action can significantly complicate the resulting performance, reflecting the distinction between what players actually do, determining payoff, and what players expect the other to do, determining action. Our simulations in Scenario \ref{chsh1} (Finding Advantage), for example, illustrate that it is possible for both players to believe in a large amount of entanglement, but nevertheless eschew actions which could leverage this resource due to having developed strong beliefs that the other would not act appropriately. An important further observation is that the random probability floor can play an instrumental role. Unexpected outcomes can cause significant belief revisions resulting in different actions being rationally preferred. Over many iterations, we see that this mechanism can produce stable behavior and performance, sometimes exceeding the classically optimal winning probability.

Evidently, expanding our considerations of quantum games in an epistemic direction is interesting, even for a game as simple as the CHSH game and a belief structure for our players which only reflects their expectations for the other's action. In the next section we will delve deeper in two ways. First, we consider an adversarial game, meaning players compete against one another, rather than a cooperative one like the CHSH game. And second, we employ a more complex belief structure for our players, explicitly endowing each agent with the belief that the other is also a rational agent. We will see that both of these modifications lead us to intriguing results.

\section{Prisoners' Dilemma}\label{sec:prisoner}
Now we turn our attention to a second game, the prisoners' dilemma. Unlike the last game, players do not simply win or lose as a team, but are instead separately rewarded, in possibly differing amounts, based on their action and the action of their opponent. This game represents the paradigmatic example of a conflict between a Pareto optimum and a Nash equilibrium. It is a ``dilemma'' because what would be best for both players if they cooperate does not align with what either can justify individually.

Investigations of iterations of this game where players' strategies can incorporate global information such as the sequence of past outcomes, an agreed upon total number of rounds, and/or their running total payoff are well known, e.g., such studies have given rise to the ``tit for tat'' strategy~\cite{Axelrod2006}. However, like the last game, we are interested in a quantum extension featuring a new single-round optimal strategy and probing whether the situation of initially uncertain players improves through sequential updates of their single-round priors. Although only focused on one round at a time, our rational agents will be deeper thinkers in a crucial way: They believe their opponent is also a rational agent and accordingly reason about the other's reasoning in a nontrivial way. 

We first review the prisoners' dilemma and the quantum extension introduced in Reference \cite{Eisert99} which brings entanglement of a state prepared by the referee, now the prison warden, into the story. After noting the optimal strategy in each case, we consider the game played by rational agents who each believes the other is rational as well, leading to an interesting simplification of the strategic landscape. Finally, we describe and run simulated iterations of this game subject to different initial priors and entanglement preparations and analyze the players' performance.

\subsection{Rules and quantum extension}
The prisoners' dilemma is a two player game, framed in terms of two imprisoned former accomplices, each separately interacting with the prison warden~\cite{Poundstone1993}. To play the game, prisoners Alice and Bob send bits $a$ and $b$ to the warden and, rather than winning or losing together, are individually rewarded according to the pair of relative values in the $a$th row and $b$th column of the payoff matrix

\begin{center}
\begin{tabular}{ c|c c } 
 (Alice, Bob) & \quad$b=0$ & \quad$b=1$\\ 
 \hline
 \rule{0pt}{3ex} $a=0$ & \quad$(3,3)$ & \quad$(0,5)$ \\ 
\rule{0pt}{3ex} $a=1$ & \quad$(5,0)$ & \quad$(1,1)$ \\ 
\end{tabular}
\end{center}
where the first number of each pair is Alice's payoff and the second is Bob's and these particular numerical values are chosen as in~\cite{Axelrod2006}. The bit value $0$ amounts to the refusal to incriminate the other prisoner while the bit value $1$ amounts to selling out their former partner. Accordingly, the two choices are referred to as ``cooperate'' (0) and ``defect'' (1). 

From the payoff matrix, one can see a few features of the game. First, mutual cooperation is said to be \emph{Pareto optimal} because no deviation from this strategy can increase the payoff for one agent without decreasing it for the other. Now consider mutual defection. This strategy is not Pareto optimal because both players would receive better payoff from mutual cooperation. However, one player unilaterally choosing to cooperate can only make their own situation worse. When no player can benefit from a unilateral deviation from a strategy, that strategy is known as a \emph{Nash equilibrium}. From the perspective of each player, defection is even stronger: each could reason that if the other cooperates, it's best for them to defect, and if the other defects, it's also best for them to defect. As defection is preferred for every choice of their opponent, it is said to be a \emph{dominant} strategy.

We can think of the payoff matrix as dictating each players' utility function. Depending on their probabilities for the action of the other, each player can compute the expected utility of any of their possible actions as well. Once we do this, we see that defection is also the rational strategy for each agent, irrespective of the details of their prior expectations. This follows simply from the fact that defection is dominant; regardless of how likely they think their opponent is to defect or cooperate, they strictly prefer to defect. This game is a dilemma because it would be better for both of them if they both cooperated, but to do so requires them both to trust each other. The issue is that the Nash equilibrium is not Pareto optimal.

Once again, if we extend the game to include shared entanglement and a larger set of possible actions, the situation of both players can improve. Reference \cite{Eisert99} introduced the quantum extension we describe here. 

Instead of directly deciding to cooperate or defect, the prisoners agree to have their play be conditioned on the following scheme. The warden sends each of them one of a pair of entangled particles upon which they act with a unitary of their choice before returning it. The warden then makes a joint measurement of the two particles, the four possible outcomes of which correspond to the four possible pairs of cooperating and defecting for the two prisoners.

Explicitly, in \cite{Eisert99}, Alice and Bob are able to implement any unitary in the two parameter family
\begin{equation}\label{unitaries}
    U(\theta,\phi)=\begin{pmatrix}
    e^{i\phi}\cos(\theta/2)&\sin(\theta/2)\\
    -\sin(\theta/2)&e^{-i\phi}\cos(\theta/2)
    \end{pmatrix}\;,
\end{equation}
with $0\leq \theta\leq \pi$ and $0\leq \phi\leq\pi/2$. 
Among these, for the purposes of faithfully containing the original game, $C:=U(0,0)=I$ and $D:=U(\pi,\cdot)=iY$ may be thought of as analogs of the cooperate and defect behaviors. 

They are each sent one half of the state 
\begin{equation}
\ket{\psi(\gamma)}=\mathcal{J}\ket{00}=\cos(\gamma/2)\ket{00}+i\sin(\gamma/2)\ket{11}\;,
\end{equation}
where $\mathcal{J}=\exp(-i\gamma D\otimes D/2)$ and, just like the CHSH game, $0\leq\gamma\leq\pi/2$ is a parameter measuring entanglement in the prepared state; when $\gamma=0$, the state is separable, and when $\gamma=\pi/2$, it is maximally entangled. After applying their unitaries and returning their qubits to the warden, the warden applies $\mathcal{J}^\dag$ to the pair and measures the result in the computational basis. Thus, for someone aware of $\gamma$ and the choices $U_A$ and $U_B$, the state before measurement is
\begin{equation}\label{finalstate}
    \mathcal{J}^\dag(U_A\otimes U_B)\mathcal{J}\ket{00}\;.
\end{equation}
Alice and Bob are rewarded according to the payoff matrix entry corresponding to the row and column specified by the two bits of the outcome of this measurement.

To assess players' potential performance, one computes the expected payoff for each agent based on the final outcome probabilities. When $\phi_A=\phi_B=0$, it is readily seen that $\gamma$ has no effect, and we may associate this entire strategic realm with the classical case of each agent independently flipping a coin with bias $\cos^2(\theta/2)$ to decide their choice. The same thing happens when $\gamma=0$; the outcome probabilities will be independent of either $\phi$ choice. However, when both are nonzero, entanglement and phase control can be utilized to strategic effect. Eisert \emph{et al.}~\cite{Eisert99} point out that when prisoners share a maximally entangled state and have the wider action space at their disposal, the action $D$ is now neither dominant nor a Nash equilibrium. Instead, a new strategy $Q:=U(0,\pi/2)=iZ$ is now a Nash equilibrium which is \emph{also} Pareto optimal with expected payoff $3$ for each player. They conclude that quantum mechanics allows prisoners to resolve the dilemma because they are no longer compelled to play a Pareto inferior strategy.\footnote{Considering this a genuine resolution of the dilemma is inappropriate because the game is changed too much by the vast expansion in action choices~\cite{vanEnk2002}. Nevertheless, it is an extension that can showcase sensitivity to entanglement and our analysis is independent of these criticisms.} 

\subsection{Belief in rationality}

If the conditions are right, quantum resources once again allow the players to better their fate, but how do a pair of rational agents fare? It turns out, as we will see, the actions that a rational player might take are greatly restricted compared to the full set of actions available in the extended game. Why should we include actions a rational agent would never play in the definition of a game? Well, just because a rational agent would never take a particular action doesn't mean they know that their opponent wouldn't take that action either. In order to know \emph{that}, the agent might further believe that their opponent is also rational like them. In the epistemic game theory literature, a rational player who thinks their opponent is also rational is called 1-fold rational~\cite{Perea2012}. Studying agents of this variety and ones with deeper layers of belief in mutual rationality (for example, a 2-fold rational player believes all other players are 1-fold rational) is central to epistemic game theory, especially when the depth is infinite and one refers to \emph{common} belief in rationality. Rather than simply considering rational agents as we did for the CHSH game, the dramatic simplification induced by 1-fold rationality motivates us to pursue the consequences of assuming Alice and Bob are agents of this type.

Consider a pair of rational agents. One may numerically verify that each agent's expected payoff is always maximized by an action with $\phi=\pi/2$ for any fixed $\gamma$ and action parameters $\theta$ and $\phi$ of their opponent. When $\theta=\pi$, this is the action $D=U(\pi,\pi/2)$, the analog of defect in the classical game, and when $\theta=0$, it is $Q=U(0,\pi/2)$, the new strategy introduced by Eisert \emph{et al.}\ which is a Pareto optimal Nash equilibrium for the maximally entangled game. Because this is true for any fixed settings of the entanglement and actions of the opponent, it will be true for any probabilistic mixture over them as well. Consequently, a rational agent will never choose a submaximal $\phi$ value. 

Remarkably, once we restrict the possible actions by setting $\phi=\pi/2$ for both agents, another simplification appears. One can explicitly show that all critical points of the resulting expected payoff landscape push rational choices to the boundaries of the parameter space, meaning, aside from a set of measure zero which we ignore, 1-fold rational agents will only ever rationally choose to play the actions corresponding to $\theta=0$ and $\theta=\pi$, that is, the actions $Q$ and $D$ and none of the actions corresponding to intermediate $\theta$ values. By the same logic as before, we are interested in what happens if we further restrict their actions to this set.

By now we have reduced the extended game back down to something much more closely resembling the original prisoners' dilemma. Like the original, our agents choose one of two actions, $Q$ and $D$. When $\gamma=0$, it is exactly the classical game with $Q$ and $D$ corresponding to cooperate and defect, regardless of the other's action. When $\gamma\neq 0$, actions and behaviors are no longer in one-to-one correspondence; if one agent plays $Q$ and the other plays $D$, these actions do not necessarily lead to the behaviors of cooperation from the first and defection from the second. In the extreme of maximal entanglement, the association is actually flipped when they take different actions --- if $\gamma=\pi/2$, Alice plays $Q$, and Bob plays $D$, the resulting behaviors are for Alice to \emph{defect} and Bob to \emph{cooperate}. Entanglement ``resolves'' the prisoners' dilemma because $\gamma$ smoothly interpolates between the prisoners' dilemma at $\gamma=0$ and a game with a dominant, Pareto optimal, Nash equilibrium at $\gamma=\pi/2$.

\begin{figure}[h]
\includegraphics[width=8cm]{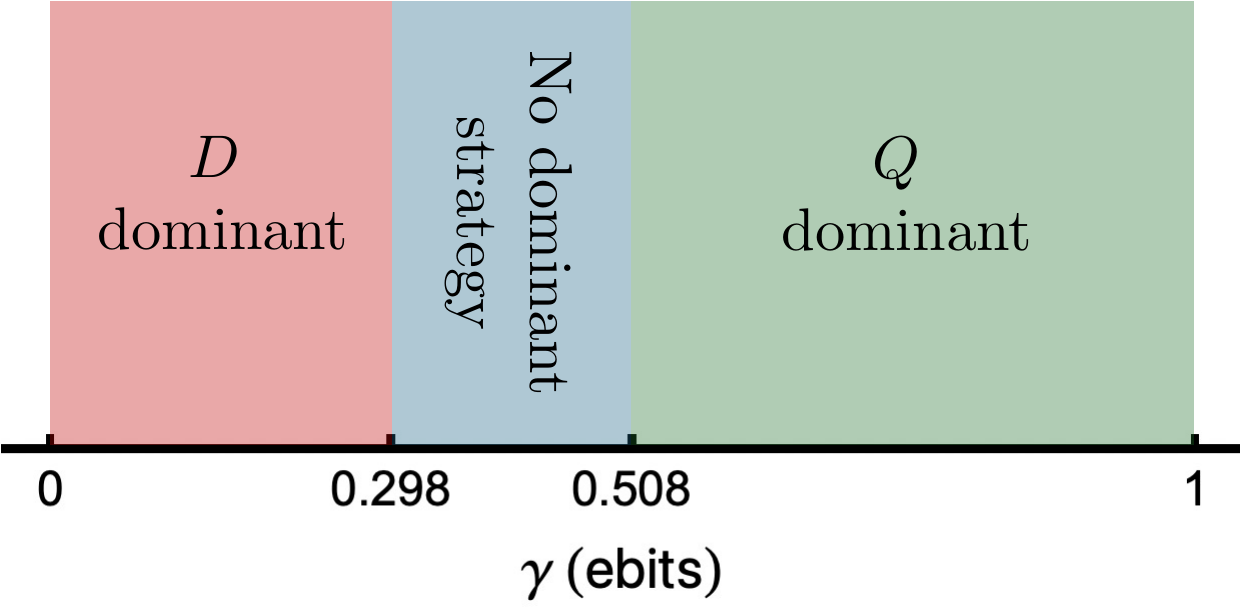}
\caption{When 1-fold rational agents play the quantum prisoners' dilemma the strategic landscape simplifies to one resembling the classical dilemma where each player chooses between two possible actions $Q$ and $D$. When the game state is separable ($\gamma=0.0$ ebits), we have the classical game with a Pareto inferior Nash equilibrium. When the game state is maximally entangled ($\gamma=1.0$ ebits), the dilemma is ``resolved'' as we have a new Nash equilibrium which is also Pareto optimal. In fact, the actions $Q$ and $D$ are dominant strategies extending into intermediate entanglement values as shown in the plot. Explicitly, when $\gamma<0.298$ ebits $D$ is dominant and when $\gamma>0.508$, $Q$ is dominant. If the game state has an entanglement in the excluded intermediate region, no action is dominant and play will explicitly also depend on what each player believes about the other's beliefs.}\label{dominantregions}
\end{figure}

One can work out that this interpolation moves through three regions with transitions at $\gamma=\arcsin(\sqrt{1/5})\approx 0.298$ ebits and $\gamma=\arcsin(\sqrt{2/5})\approx 0.508$ ebits, see Figure \ref{dominantregions}. When $\gamma<\arcsin(\sqrt{1/5})$, $D$ is a dominant strategy and when $\gamma>\arcsin(\sqrt{2/5})$, $Q$ is dominant.  In between these values there is no dominant action and so, for rational agents, what they play will depend not only on their beliefs about the game state, but also their beliefs about what their opponent will do. For 1-fold rational agents, what each believes their opponent will do in turn depends on what they believe their opponent believes.

Because they are 1-fold rational, each agent considers the other to be rational agents, that is, agents who possess beliefs about the entanglement of the game state and beliefs about what actions their opponent will take. As they believe their opponent is rational, they know their opponent also reaches a choice by maximizing expected utility. Thus, in addition to her own distribution over possible game state entanglements, Alice, for example, possesses a prior for Bob's beliefs about the game state entanglement and for his beliefs about \emph{Alice's} action. Note that Alice \emph{does not} have beliefs about what she thinks Bob's beliefs about her beliefs are --- this would be a 2-fold rational agent, namely, a rational agent who believes their opponent is a 1-fold rational agent.

To make her decision, she first computes a probability that Bob will take each of the two possible actions by considering what his rational choice would be for each combination of possible beliefs she has about his beliefs. With this and her entanglement prior, she arrives at a rational action. Bob does the same. After each receives an outcome, each updates their entanglement prior with Bayes rule and updates their prior about the others' beliefs by computing the Bayes rule posterior that an agent possessing those beliefs would have upon receiving the appropriate outcome. Once this is complete, the game can repeat. 

\subsection{Simulated iterations}

In this subsection we describe and discuss some simple simulations of repeated rounds of 1-fold rational agents playing the prisoners' dilemma game starting from particular priors and game state preparations and where the state of knowledge of each player changes according to Bayes rule between each round. 

Our purpose for running these simulations is to probe features of performance and belief dynamics going beyond what we have already seen with the CHSH game. In particular, since the players now have beliefs about what the other believes, we can explore the consequences for game performance of a player holding \emph{incorrect} beliefs about the other's beliefs. Similarly, the structure of this game permits players' beliefs about the presence or absence of a resource to have a greater impact on performance allowing us, for example, to ask how different mutual belief in entanglement is from its actual presence? 

\subsubsection{Discretizations and prior structure}

To simulate 1-fold rational agents playing the prisoners' dilemma, we again impose discretizations and a quantum probability floor as in our simulations of the CHSH game described in \S\ref{chshdiscretizations}. 

The game state entanglement is discretized into the same $0.1$ ebit values as before and each agent again has a prior over the possible set of entanglement values. Now, however, instead of a prior directly over the possible actions of their opponent, each agent should have a prior over the possible entanglement priors they believe their opponent could believe and over a set of possible biases towards $D$ or $Q$ that they believe their opponent believes about them. We make a slightly unrealistic assumption regarding beliefs about entanglement beliefs to simplify simulations which retains the crucial element of beliefs about beliefs: Rather than having to work with a distribution over distributions, we assume each believes their opponent is certain the game state is one of the 11 possible choices, and so their beliefs about their opponent's entanglement beliefs is a probability over this set. Finally, we discretize the possible biases to be the 11 values between 0 and 1 in steps of 0.1, where the value 1 corresponds to certainty of $D$ and 0 corresponds to certainty of $Q$. 

Thus, Alice has a probability distribution over the 11 possible biases she believes Bob can believe about the action Alice will take as well as a distribution over the 11 possible entanglement values Bob could (with certainty) believe. Similarly for Bob. Overall, each has a joint discrete probability distribution over the game state entanglement and over the possible entanglement and biases of their opponent. 

We again implement a probability floor to mitigate the damage of certainty (see \S\ref{sec:rationalplayers}) and we use essentially the same protocol as before, but now for the two qubit computational basis measurement that the referee will make to conclude each round. Explicitly, we make the substitution $P\rightarrow (1-\epsilon)P+\frac{\epsilon}{4}I$ with $\epsilon=0.1$ for each of the four projectors.

\subsubsection{Simulation scenarios}\label{prisonerScenarios}

\begin{figure*}
     \begin{subfigure}[t]{0.49\textwidth}
         \includegraphics[width=\textwidth]{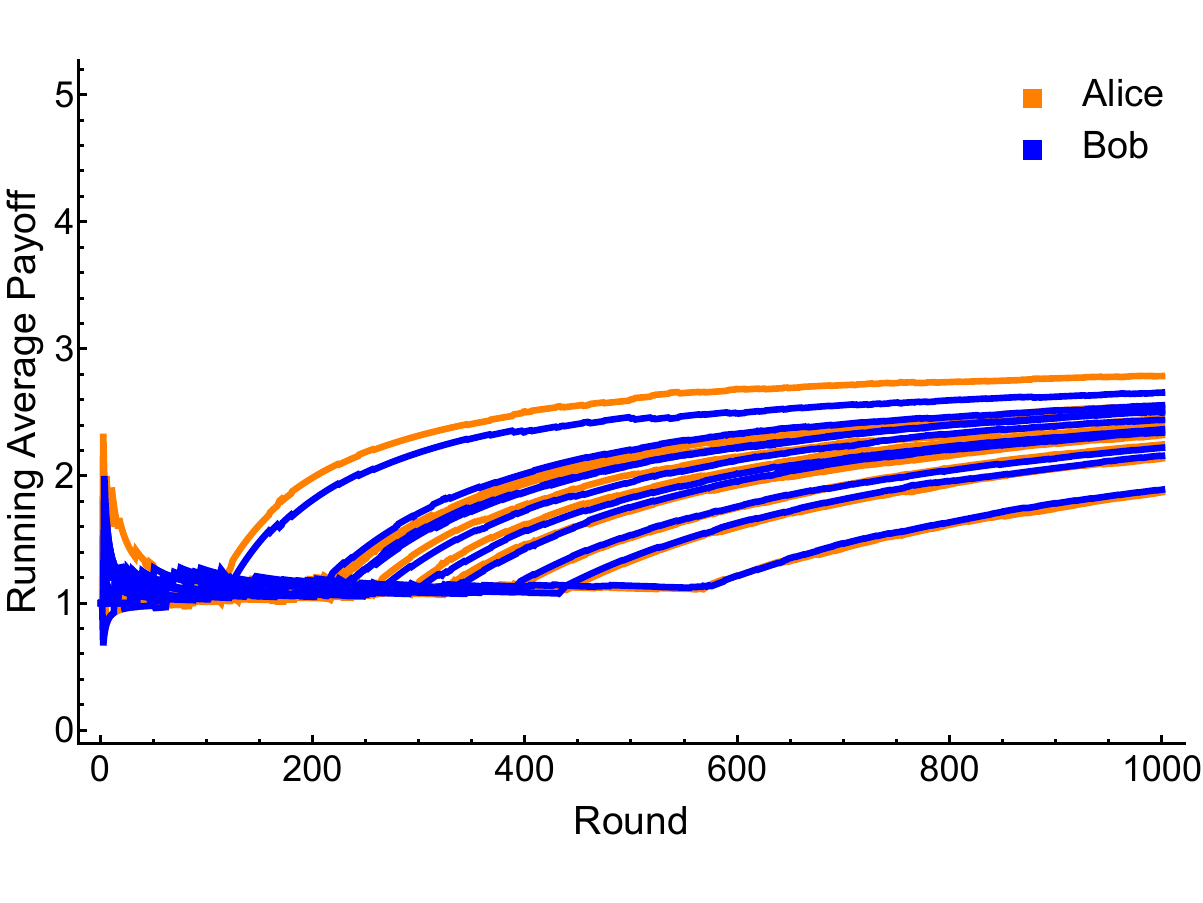}
         \caption{Prisoner Scenario \ref{prisoner1} (Bohr's Horseshoe): $\gamma_g=1.0$ ebits, Alice (orange) and Bob (blue) both start with the \hyperlink{lowlow}{\color{black}low-low} prior. Initially both play $D$ and defect with high probability. The probability floor leads one or the other to occasionally cooperate, shifting entanglement beliefs. Eventually, after an average of 303 rounds, one of them expects entanglement in the $Q$ dominant region and the subsequent outcomes drive up the other's entanglement expectations. Both play $Q$ to mutual cooperation thereafter.}
         \label{subfig:pris1}
     \end{subfigure}
     \hfill
     \begin{subfigure}[t]{0.49\textwidth}
         \includegraphics[width=\textwidth]{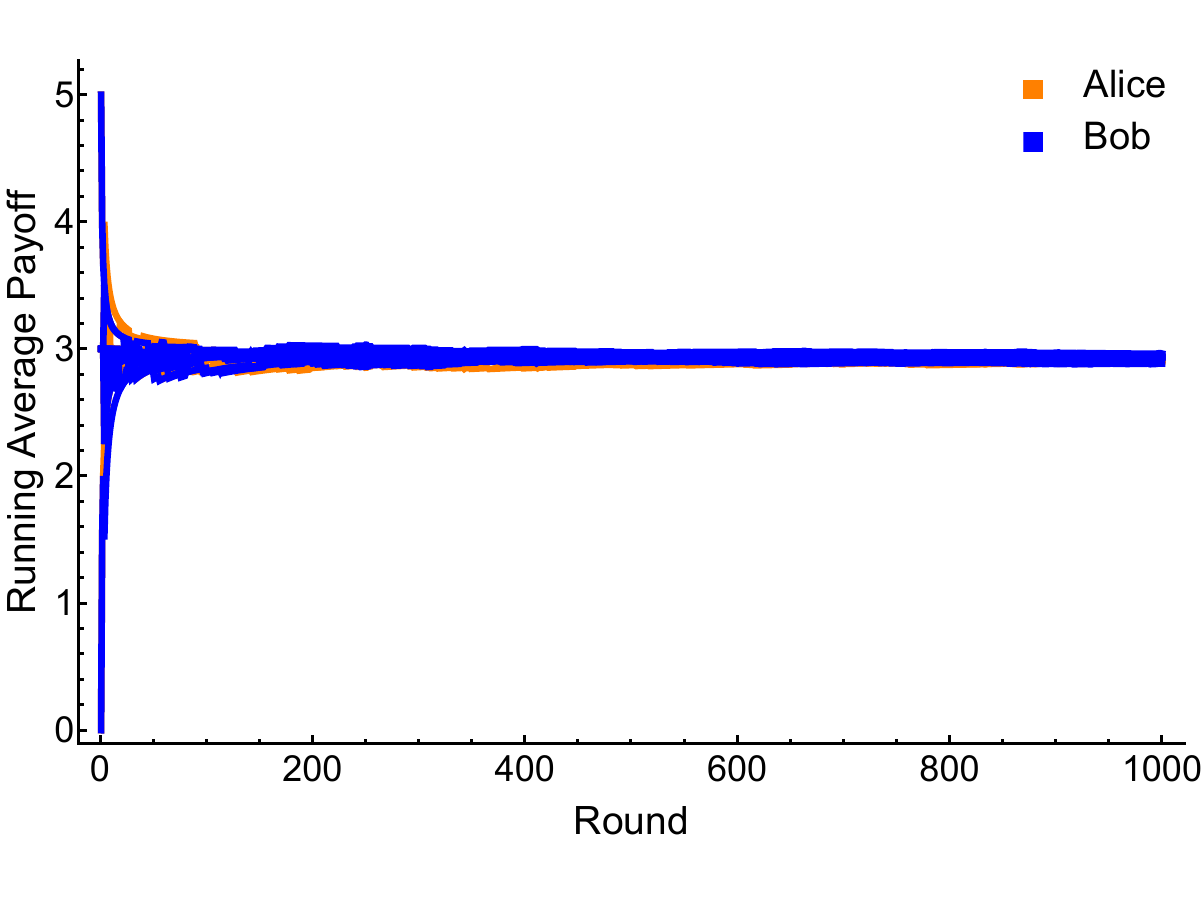}
         \caption{Prisoner Scenario \ref{prisoner2} (Faith Alone): $\gamma_g=0.0$ ebits, Alice (orange) and Bob (blue) both start with the \hyperlink{highhigh}{\color{black}high-high} prior. Although the game state is separable, their belief in high entanglement leads them play $Q$ and cooperate with high probability in every round. When one defects, entanglement priors shift, but this does not prove to be enough to change their behavior in the number of rounds simulated.}
         \label{subfig:pris2}
     \end{subfigure}
     \hfill
     \begin{subfigure}[t]{0.49\textwidth}
         
         \includegraphics[width=\textwidth]{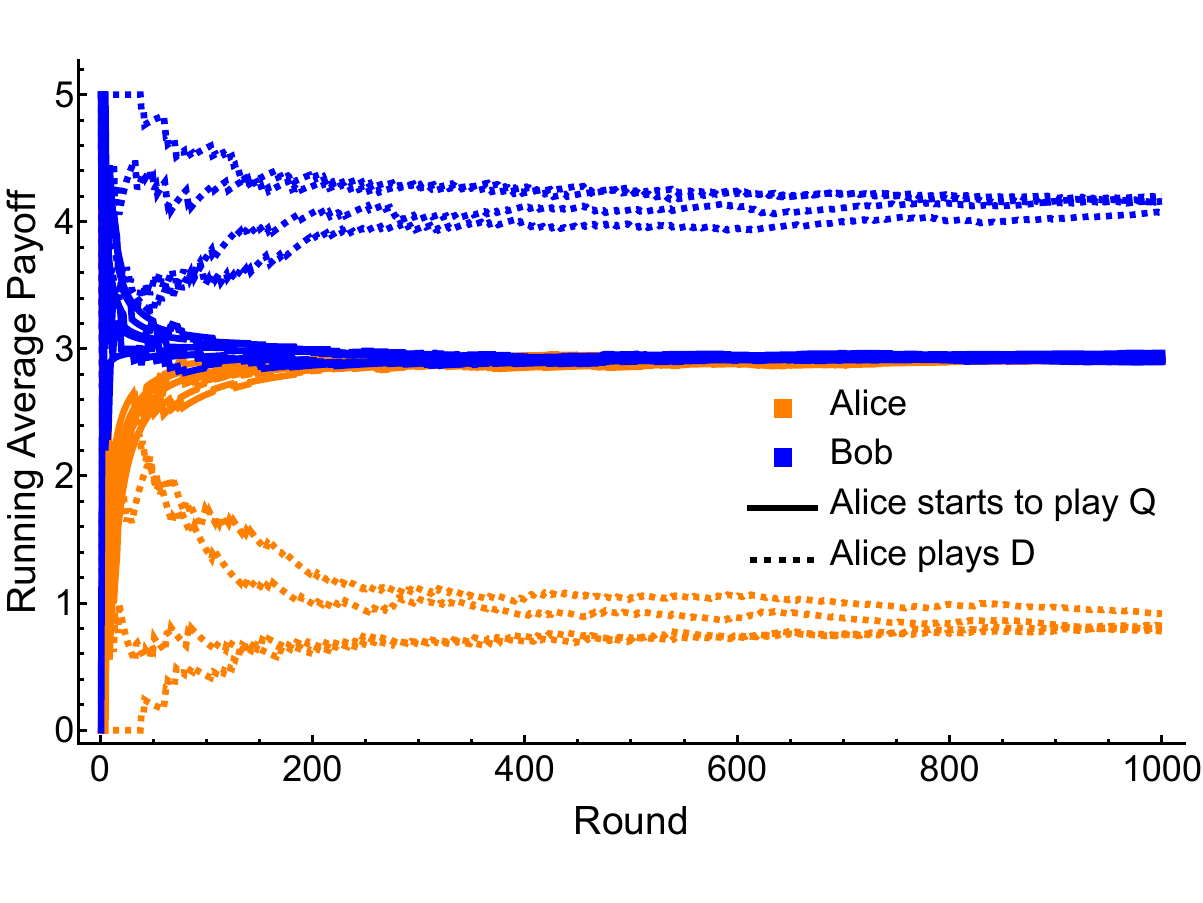}
         \caption{Prisoner Scenario \ref{prisoner3} (Double Down?): $\gamma_g=0.9$ ebits, Alice (orange) starts with the \hyperlink{lowlow}{\color{black}low-low} prior and Bob (blue) starts with the \hyperlink{highhigh}{\color{black}high-high} prior. Initially Alice plays $D$ and Bob plays $Q$. As the game state is highly entangled, the most likely outcome is for Alice to cooperate and Bob to defect, which pushes Alice's entanglement expectation up into the region without a dominant strategy. Updating to higher entanglement expectation competes with Alice's confidence that Bob will play $D$ until one action for Alice dominates. $Q$ wins in the six solid line simulations, leading to mutual cooperation thereafter. $D$ wins in the four dotted simulations, leading to her cooperation and Bob's defection thereafter.}
         \label{subfig:pris3}
     \end{subfigure}
     \hfill
     \begin{subfigure}[t]{0.49\textwidth}
        
         \includegraphics[width=\textwidth]{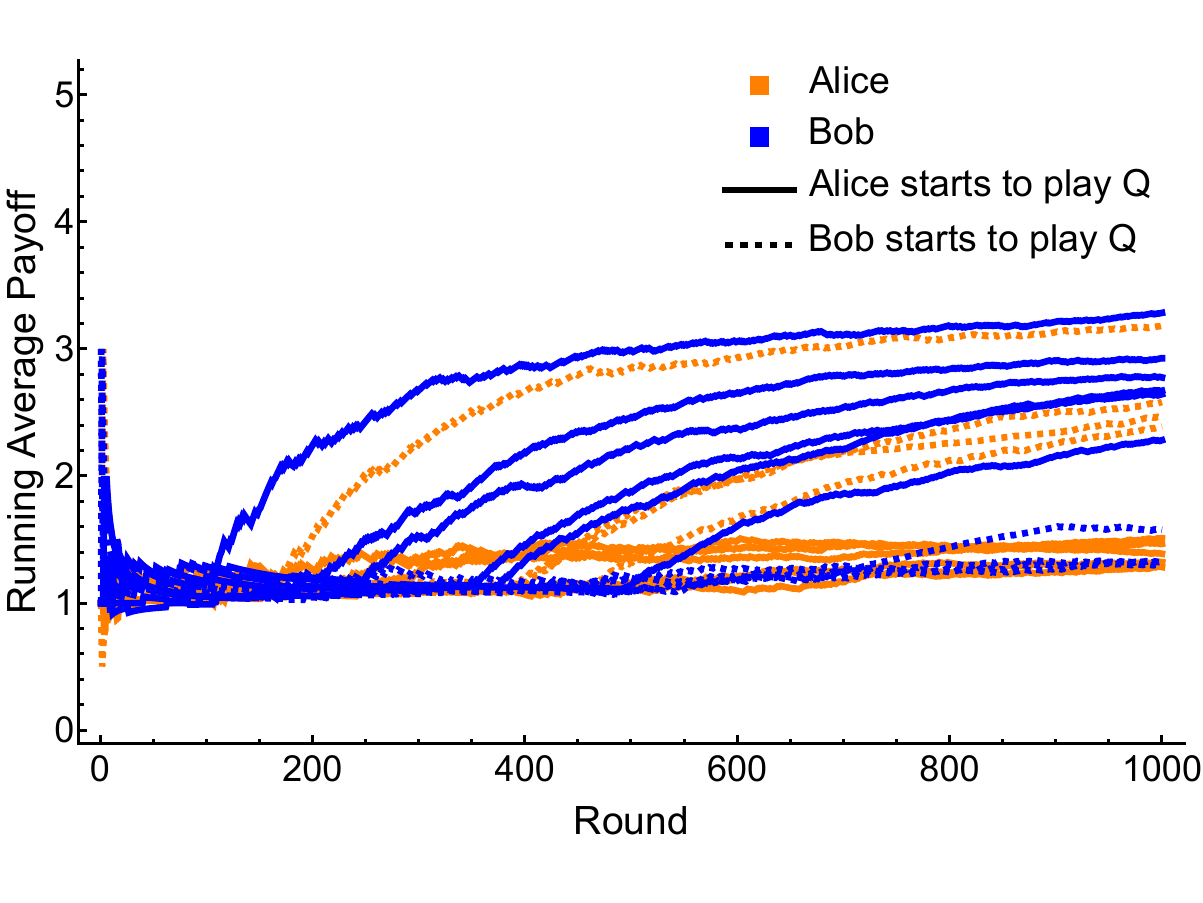}
         \caption{Prisoner Scenario \ref{prisoner4} (Fool's Gold): $\gamma_g=0.4$ ebits, Alice (orange) and Bob (blue) both start with the \hyperlink{lowhigh}{\color{black}low-high} prior. Initially both play $D$ and defect with high probability, although both expect the other to play $Q$. When probability floor events push one of their entanglement expectations into the $Q$ dominant region, they have also become essentially certain the other will play $Q$, when, in actuality, the other continues playing $D$, typically leading to a worse outcome for the first. Alice switches to playing $Q$ in the six solid line simulations while Bob switches in the remaining 4 dotted line simulations. Overall, one of them switches after an average of 320 rounds.}
         \label{subfig:pris4}
     \end{subfigure}
        \caption{Cumulative average payoff for simulations of 1-fold rational agents Alice (orange) and Bob (blue) playing 10 simulations of 1000-rounds of the quantum prisoners' dilemma for each of the four prior and game state scenarios described and analyzed in \S\ref{prisonerScenarios}.  After each round, both players update their single-round priors with Bayes rule before the next round begins.}
        \label{fig:prisonerPayoff}
\end{figure*}

\begin{figure*}
     \begin{subfigure}[t]{0.49\textwidth}
         \centering
         \includegraphics[width=\textwidth]{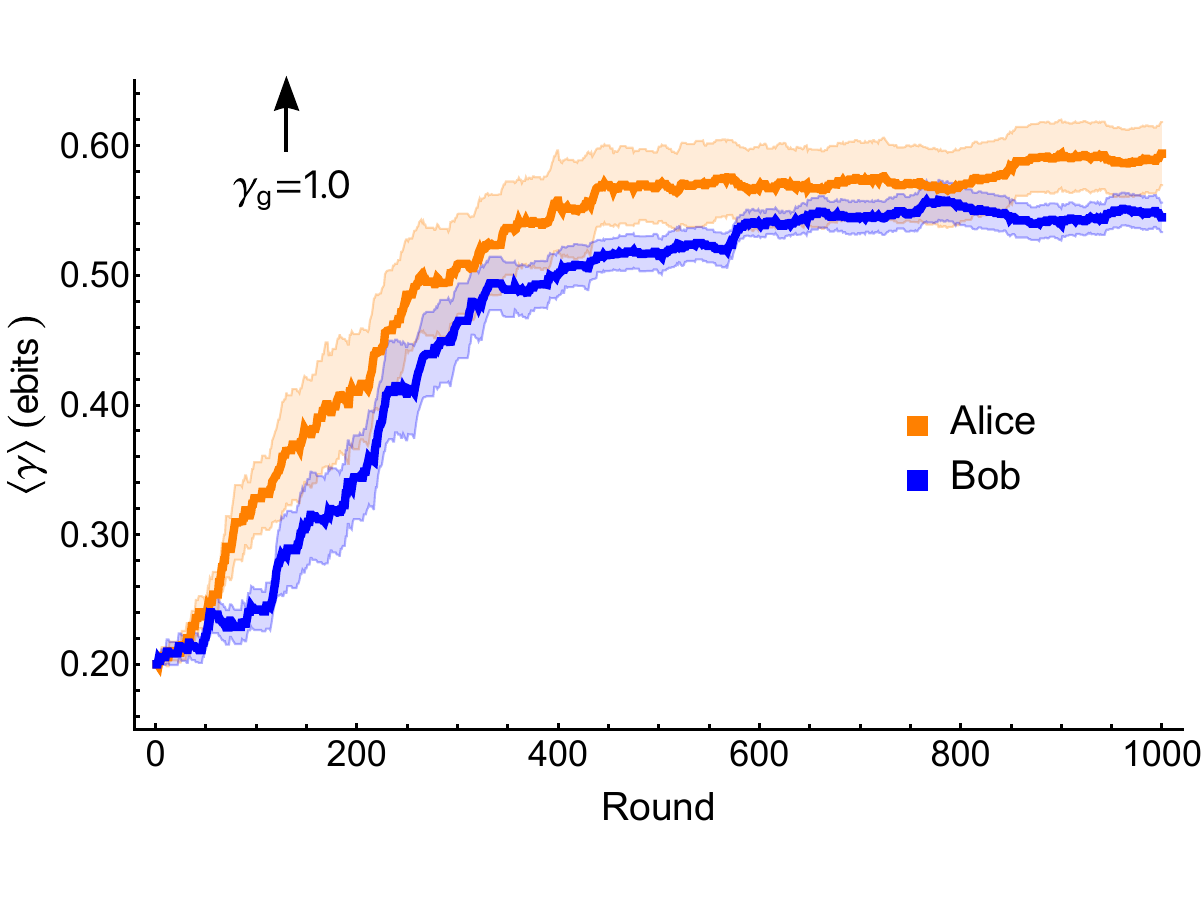}
         \caption{Prisoner Scenario \ref{prisoner1} (Bohr's Horseshoe): $\gamma_g=1.0$ ebits (above plotted range), Alice (orange) and Bob (blue) both start with the \hyperlink{lowlow}{\color{black}low-low} prior. Initially both play $D$ and defect with high probability. Probability floor events occasionally cause one of them to cooperate, driving up their entanglement expectation and the other's down. After an average of $303$ rounds, one of the players enters the $Q$ dominant region and subsequent outcomes drive the other's into this region as well. Once mutually cooperating, outcomes do not depend on game state so their expectations level off.}
         \label{subfig:prisENT1}
     \end{subfigure}
     \hfill
     \begin{subfigure}[t]{0.49\textwidth}
         \includegraphics[width=\textwidth]{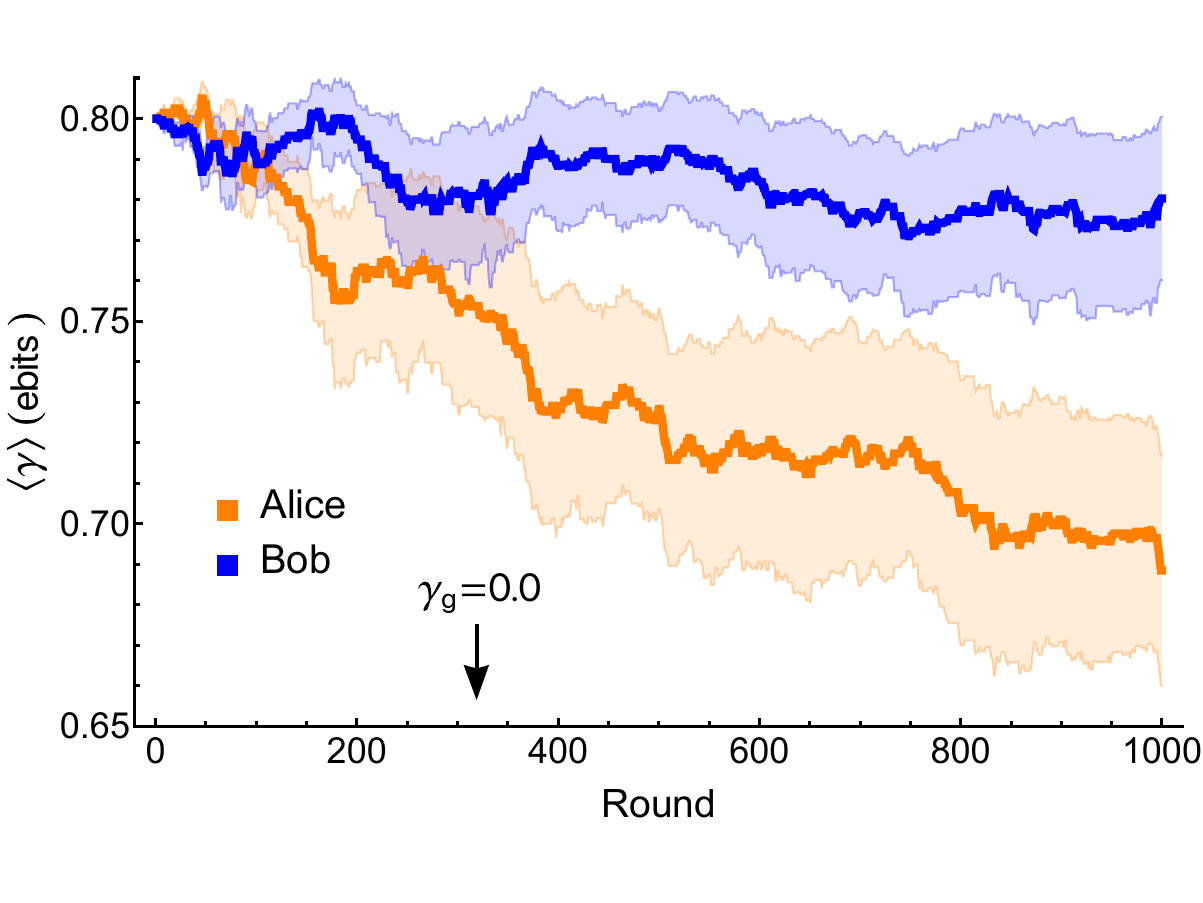}
         \caption{Prisoner Scenario \ref{prisoner2} (Faith Alone): $\gamma_g=0.0$ ebits (below plotted range), Alice (orange) and Bob (blue) both start with the \hyperlink{highhigh}{\color{black}high-high} prior. Each plays $Q$ and expects the same of their opponent. Their outcomes are in fact insensitive to the game state, but when a probability floor event causes one of them to cooperate, that agent's entanglement prior shifts reflecting the possibility their opponent played $D$ in a lower entanglement game. In our simulations this shift happened more to Alice than to Bob, but this does not reflect a systematic bias.}
         \label{subfig:prisENT2}
     \end{subfigure}
     \hfill
     \begin{subfigure}[t]{0.49\textwidth}
         \includegraphics[width=\textwidth]{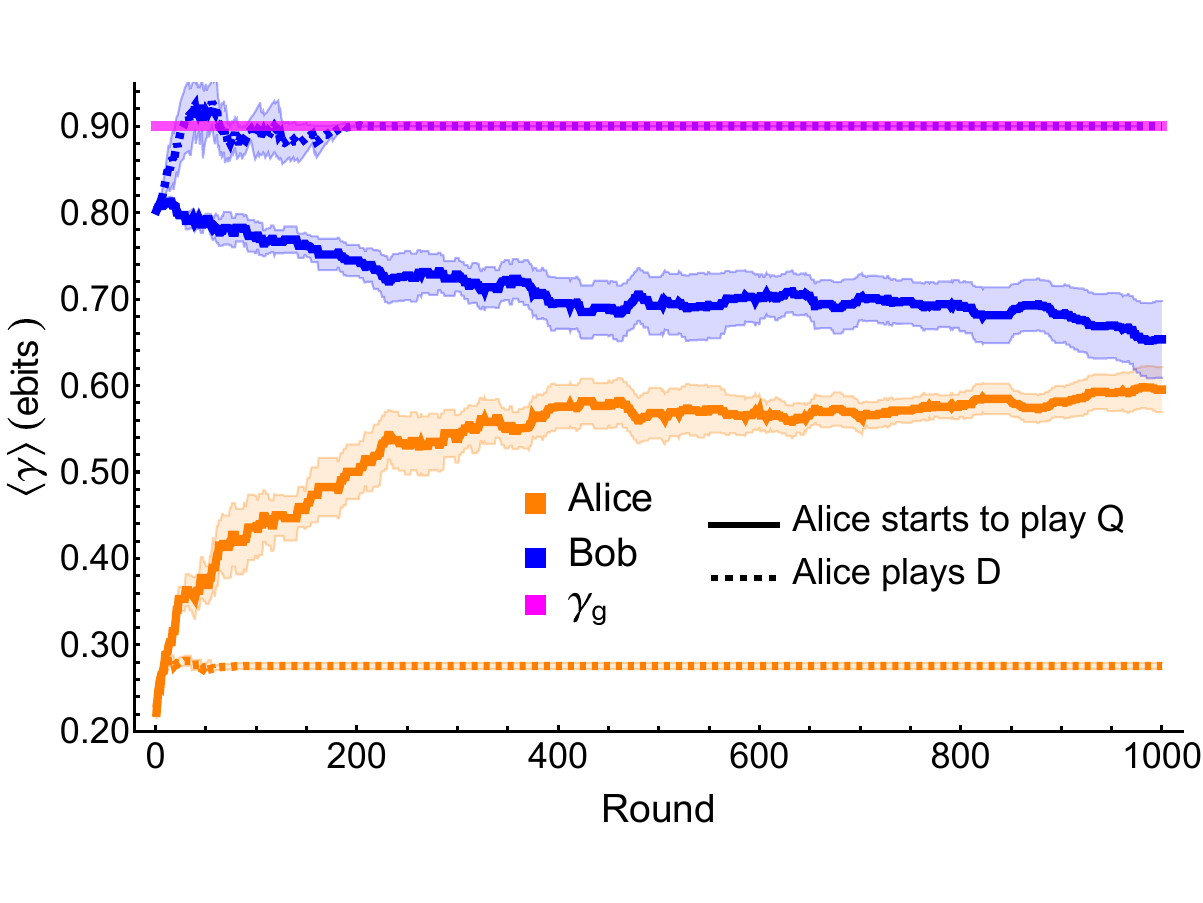}
         \caption{Prisoner Scenario \ref{prisoner3} (Double Down?): $\gamma_g=0.9$ ebits (pink), Alice (orange) starts with the \hyperlink{lowlow}{\color{black}low-low} prior and Bob (blue) starts with the \hyperlink{highhigh}{\color{black}high-high} prior. The solid lines are the average and standard deviation of the expected entanglement in the 6 simulations where Alice switches to playing $Q$ and the dotted lines are the average and standard deviation of the expected entanglement in the 4 simulations where she settles on $D$. Initially Alice plays $D$ and Bob plays $Q$. When Alice switches to $Q$, subsequent outcomes are insensitive to the game state, but probability floor events shift each player's prior in opposite directions. When Alice settles on $D$, her entanglement prior freezes while Bob's entanglement expectations are refined to the game state entanglement within about 100 rounds.}
         \label{subfig:prisENT3}
     \end{subfigure}
     \hfill
     \begin{subfigure}[t]{0.49\textwidth}
         \includegraphics[width=\textwidth]{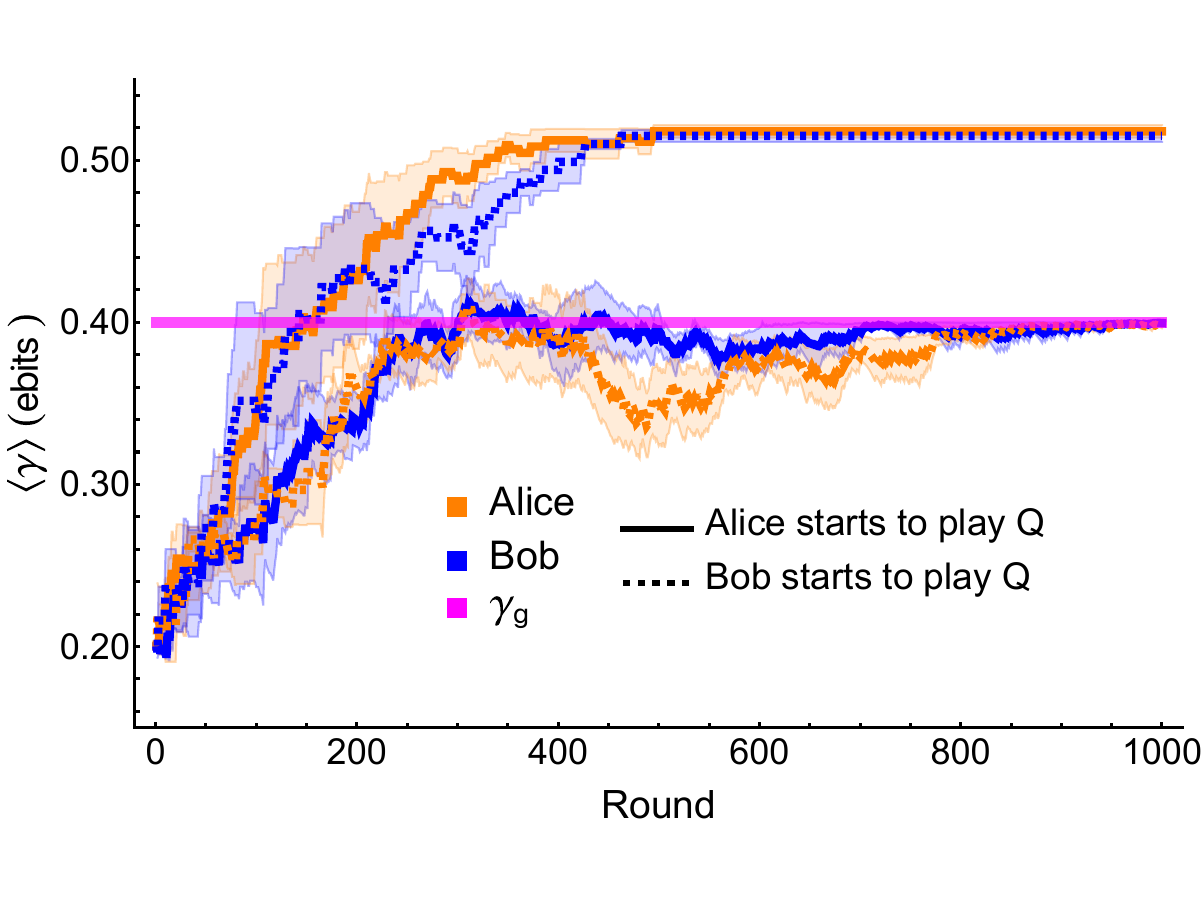}
         \caption{Prisoner Scenario \ref{prisoner4} (Fool's Gold): $\gamma_g=0.4$ ebits (pink), Alice (orange) and Bob (blue) both start with the \hyperlink{lowhigh}{\color{black}low-high} prior. The solid lines plot the average and standard deviation of the expected entanglement in the 6 simulations where Alice starts to play $Q$ and the dotted lines plot the average and standard deviation of the expected entanglement in the the 4 simulations where Bob starts to play $Q$. Initially both play $D$ and defect with high probability, although both expect the other to play $Q$. Probability floor events eventually drive one of them into the $Q$ dominant region and makes them essentially certain the other will also play $Q$, freezing their entanglement expectations. Once this happens, the other’s expectation that the first will play Q is correct and subsequent outcome frequencies drive their entanglement expectation to the game state value.}
         \label{subfig:prisENT4}
     \end{subfigure}
        \caption{Expected entanglement plots for simulations of 1-fold rational agents Alice (orange) and Bob (blue) playing 10 simulations of 1000-rounds of the quantum prisoners' dilemma for each of the four prior and game state scenarios described and analyzed in \S\ref{prisonerScenarios}. Bands around lines indicate the average and standard deviation. After each round, both players update their single-round priors with Bayes rule before the next round begins.}
        \label{fig:prisonerEntanglement}
\end{figure*}

As for the CHSH game, we consider four scenarios of the quantum prisoners' dilemma, varying players' priors and the game state entanglement, which we call ``Bohr's Horseshoe'', ``Faith Alone'', ``Double Down?'', and ``Fool's Gold''. For each scenario we ran 10 simulations of 1000 rounds. We summarize our results in Figures \ref{fig:prisonerPayoff} and \ref{fig:prisonerEntanglement}. Figure \ref{fig:prisonerPayoff} tracks, for each of the 10 simulations of each scenario, the cumulative running average payoff for each player as the rounds progress and Figure \ref{fig:prisonerEntanglement} tracks how players' entanglement expectations evolve. 

In the scenarios below we initialize each agent with one of three initial priors. For all three, the probability distribution over the other's biases for the first's action is uniform, so these priors differ only in the entanglement beliefs of oneself and one's opponent. A ``low'' entanglement prior for us is, explicitly, the binomial distribution for the number of successes in 10 attempts for an event with probability $1/5$ and a ``high'' entanglement prior is the same, but for an event with probability $4/5$. Then the initial priors we consider are as follows: 
\begin{lowlow}\hypertarget{lowlow}The agent believes the game state has low entanglement and believes their opponent also believes the game state has low entanglement. Their distribution over the other's action biases is uniform.
\end{lowlow}
\begin{highhigh}\hypertarget{highhigh} The agent believes the game state has high entanglement and believes their opponent also believes the game state has high entanglement. Their distribution over the other's action biases is uniform.
\end{highhigh}
\begin{lowhigh}\hypertarget{lowhigh}The agent believes the game state has low entanglement, but believes their opponent believes the game state has high entanglement. Their distribution over the other's action biases is uniform.
\end{lowhigh}

\begin{prisoner}[Bohr's Horseshoe]\label{prisoner1} The warden prepares the maximally entangled game state, $\gamma_g=1.0$ ebits, and Alice and Bob both have the \hyperlink{lowlow}{\color{black}low-low} initial prior. A quantum resource, entanglement, is present, but nobody believes in it. We will see this is an instance of Bohr's Horseshoe\footnote{This is a reference to the popular anecdote that Niels Bohr kept a horseshoe over a doorway in his house, not because he believed it brought luck, but because he heard it works even if you don't believe. Regarding its authenticity, see \cite{Stacey2011}}: Entanglement helps the players even though they don't (initially) believe in it.

Figure \ref{subfig:pris1} tracks the cumulative running average payoff for this scenario. With low entanglement expectation, the initial rational action for both players is to play $D$. This leads both to defect with high probability, the Nash equilibrium of the classical game. This outcome teaches them nothing about the entanglement, but the probability floor results in one or the other occasionally cooperating and leading to a shift in their entanglement priors. Eventually, after an average of 303 rounds, one of them expects an entanglement value in the $Q$ dominant region and begins playing $Q$ every time. The other quickly catches on as they find themselves cooperating despite playing $D$. This outcome is most likely if the other were to play $Q$ in the presence of higher entanglement; accordingly, Bayes rule drives up their entanglement expectation into the dominant region as well after an average of 10 rounds. For the remaining rounds, both play $Q$ and mutually cooperate.

Figure \ref{subfig:prisENT1} shows how Alice and Bob's entanglement expectations evolve. One sees that both agent's entanglement beliefs rise towards the game state entanglement but then level off. When probability floor events push one into the $Q$ dominant region, there is a period of quick entanglement expectation growth for both of them with each round. Once the other plays $Q$ as well, their outcomes are insensitive to the game state and they return to only having nontrivial entanglement prior updates when rare probability floor events occur. As this effect equally leads to shifts in both directions, there is predominantly a leveling off of entanglement expectations the in later rounds.
\end{prisoner}

\begin{prisoner}[Faith Alone]\label{prisoner2} The warden prepares the separable game state, $\gamma_g=0.0$ ebits, and Alice and Bob both have the \hyperlink{highhigh}{\color{black}high-high} initial prior. Will their faith alone help them?

Figure \ref{subfig:pris2} tracks the cumulative runnning average payoff for this scenario. The behavior of Alice and Bob is to mutually cooperate for every round of every simulation modulo the probability floor. Indeed, since they each believe in high enough game state entanglement, each plays $Q$ and cooperates with high probability in every round. Running average payoff is reduced slightly by the probability floor events; in fact, in 2 simulations the probability floor caused one agent to defect in the first round leading to the initial payoff spread.

While the probability floor events do not prove to be enough to change their behavior over the timescale of the simulations, it appears as though it has an effect on the players' expected entanglement as can be seen in Figure \ref{subfig:prisENT2}. Here it is important to remember that the Alice and Bob's situation is symmetric in this scenario and so the distinction between their entanglement expectations is due to sampling rather than a systematic bias. In fact, as they both play $Q$ in every round, the outcomes they receive are entirely independent of the game state. As entanglement beliefs are driven down by probability floor events, the fact that the game state is separable stands a chance of destabilizing the players' behavior, but this possibility must be weighed against a potential increasing certainty of their beliefs about their opponent.

While a simple example, it is notable that belief in entanglement alone is enough in this case for rational agents to sidestep the original dilemma. Particularly, comparing to Scenario \ref{prisoner1} (Bohr's Horseshoe), we see that believing in entanglement that is not there is better for the players than disbelieving in entanglement that is there! The players' belief in (nonexistent) entanglement is equivalent to any belief that the other player will always cooperate: it is a proxy for trust.
\end{prisoner} 

\begin{prisoner}[Double Down?]\label{prisoner3} The warden prepares the $0.9$ ebit game state, $\gamma_g=0.9$ ebits, Alice has the \hyperlink{lowlow}{\color{black}low-low} initial prior, and Bob has the \hyperlink{highhigh}{\color{black}high-high} initial prior. Will Alice realize she was wrong about the entanglement and Bob's beliefs or will she double down?

Figure \ref{subfig:pris3} tracks the cumulative running average payoff for this scenario. Initially, Alice plays $D$ and Bob plays $Q$. As the game state is highly entangled, the most likely outcome is for Alice to cooperate and Bob to defect. For Alice, given she played $D$, this outcome is either a probability floor event or a consequence of Bob playing $Q$ on a high entanglement game state. However, she believes the game state entanglement is low and she also believes that Bob thinks so as well, meaning he is highly likely to play $D$. Updating her own prior to higher entanglement thus competes with her confidence that Bob will play $D$, which itself varies with each round. For Bob, given he played $Q$, the most likely outcome of Alice cooperating and his defection is similarly either a probability floor event or a consequence of Alice playing $D$ on a high entanglement game state. He already believes in high entanglement himself so events like this are not as surprising to Bob as they are for Alice and, accordingly, her expectations move more.

This effect can shift Alice out of the $D$ dominant region and into one that depends crucially on her beliefs about Bob's beliefs, see Figure \ref{dominantregions}. This happened in 8 of our simulations; in the other 2, Alice became overwhelmingly confident that Bob will play $D$, solidifying her initial choice to play $D$ herself. Of the 8 simulations which came to depend on her beliefs about Bob, in 2 more she nevertheless ends up overwhelmingly certain that Bob will play $D$, causing her to settle back to $D$ after a short stint playing $Q$. In the other 6 simulations, however, she loses confidence that Bob will play $D$ and her entanglement expectation grows substantially enough that she begins to play $Q$, leading to mutual cooperation thereafter. 

Figure \ref{subfig:prisENT3} presents the entanglement expectations for both players in this scenario. If Alice switches to $Q$, both players play $Q$ every time and their outcomes simply do not depend on $\gamma_g$. This is the same situation as in Scenario \ref{prisoner1} (Bohr's Horseshoe) except this time Bob's entanglement expectation starts high. The apparent attraction of their expectations is due to the fact Alice and Bob's entanglement expectations will shift in opposite directions for a given probability floor event of one of them defecting. When Alice fails to switch from $D$, growing ever more certain that Bob will also play $D$, her entanglement expectation freezes at an average of $0.28$ ebits. Over an average of 150 subsequent rounds, Bob's entanglement beliefs become completely localized exactly at the game state value. This works because, up to probability floor deviations, he will be receiving the outcome he most expects, Alice's cooperation and his defection, at precisely the frequency that a game state of $0.9$ ebits should produce. 
\end{prisoner} 

\begin{prisoner}[Fool's Gold]\label{prisoner4} The warden prepares the $0.4$ ebit game state, $\gamma_g=0.4$ ebits, and Alice and Bob both have the \hyperlink{lowhigh}{\color{black}low-high} initial prior. Both are wrong about what the other thinks. As they come to believe in more entanglement, playing $Q$ can become more attractive even though it will actually worsen the relative situation of whomever switches! 

Figure \ref{subfig:pris4} tracks the cumulative running average payoff for this scenario. Initially, Alice and Bob both play $D$ and defect with high probability, although both expect the other to play $Q$. Probability floor events push them out of the $D$ dominant region and into the indeterminate region, as in the previous scenario, but this time, because they believe the other \emph{wrongly} has a high entanglement prior, their confidence that the other will play $Q$ inspires them to continue playing $D$. Only when probability floor events manage to raise one of their entanglement expectations into the $Q$ dominant region does something different happen. By then, it turns out, they have typically also become essentially certain that the other will play $Q$, when, in actuality, the other continues playing $D$. The one who switched to $Q$, thanks to the  certainty they have developed, regards the outcome of their cooperation and the other's defection as more likely to have resulted from a probability floor event than from the other playing $D$, when, in actuality, the other \emph{did} play $D$ and this outcome was the most likely one. In 6 simulations, Alice starts to play $Q$ and in the other $4$, Bob does. 

Figure \ref{subfig:prisENT4} presents the entanglement expectations for both players in this scenario. Once one of the players has switched to $Q$ and believes essentially with certainty that the other will play $Q$ as well, he or she fails to learn further about the game state entanglement as in Scenario \ref{prisoner3} (Double Down?) when Alice settled on $D$. This results in a strongly peaked entanglement prior at an average of about $0.52$ ebits. Once again, the agent that did not switch subsequently detects the game state entanglement perfectly. Over the course of an average of almost 500 further rounds, the other agent sharpens their beliefs until they are extremely close to the correct game state entanglement value.
\end{prisoner}

\section{Summary and Outlook}\label{sec:discussion}

We have studied Bayesian rational agents playing iterations of quantum games. Most broadly, we have seen that explicitly considering players' beliefs in this setting is interesting and subtle. The usual point of constructing quantum games is to show the possibility for a quantum advantage over games with only classical resources. We promoted our players to genuine decision-makers for two reasons. The first was to see whether agents who know a game's setup and quantum mechanics are generically able to find a quantum advantage. The second was taken from QBism and epistemic game theory which both demand that reasoning and agency be taken seriously: How does including the knowledge and beliefs of players in our study of quantum games change and add to the phenomena we can observe? 

To investigate these, we examined and simulated iterations of two well-known quantum games, the CHSH game and the quantum prisoners' dilemma. In both games, when outcomes are generated from an entangled state, initially ignorant rational agents who possess single-round priors can often learn through iterated play to find actions which are superior to the best possible classical strategies. In the process, we find that players beliefs about entanglement will typically shift towards the game state preparation of the referee or warden as long as their payoff is dependent on this entanglement. 

Far from being an uninteresting complication, we find that explicitly modeling players' beliefs can result in performance which intricately depends on players' beliefs about their opponents and their beliefs about the structure of the game. As it is belief, not truth, which guides action, we encounter situations where belief in the presence or absence of a quantum resource matters as much or more than the fact of the matter (See Prisoner Scenarios \ref{prisoner1} (Bohr's Horseshoe) and \ref{prisoner2} (Faith Alone)) or where a player tragically becomes nearly certain, yet wrong, about the action or beliefs of their opponent leading them to rationally prefer suboptimal actions (See the discussion of CHSH Scenario \ref{chsh1} (Finding Advantage) and Prisoner Scenarios \ref{prisoner3} (Double Down?) and \ref{prisoner4} (Fool's Gold)).

The most dramatic consequence we found of modeling rational agents occurred when considering the prisoners' dilemma, where
we considered rational players who also explicitly believe in the rationality of their opponent. This revealed an interesting effect which structured our simulations: For agents of this level of sophistication, generalizing the classical game to a continuous quantum strategic landscape is inappropriate. Instead we end up with a single entanglement parameter which interpolates between the classical game when separable to another discrete game with a Pareto optimal Nash equilibrium for large enough entanglement. Accounting for players' beliefs about the other's beliefs simplifies the game structure while producing highly nontrivial behavior such as that seen in Prisoner Scenarios \ref{prisoner3} (Double Down?) and \ref{prisoner4} (Fool's Gold).

Prisoner Scenario \ref{prisoner2} (Faith Alone) also represents a straightforward yet powerful example of the power of beliefs. Both Alice and Bob believe the game state is highly entangled and also believe the other believes this as well, yet, in actuality, the warden prepares a separable state. Due to the game's structure, belief in entanglement alone resolves the dilemma: The players mutually cooperate for the duration of the rounds. Of course, belief in entanglement is not a long-term substitute for its presence as probability floor events will eventually chip away at the players' entanglement expectations, in other words, mutual cooperation is ultimately unstable to fluctuations, but this does not much trouble our Alice and Bob whose beliefs alone guided them through a thousand rounds of mutual cooperation.

The numerical aspects of our work should be regarded primarily as proofs of principle. In particular, there are many places where different choices could have been made regarding what to simulate and precisely how. For example, which elements of gameplay we regard as visible to a player will fundamentally alter how they act and how they learn from outcomes. Similarly, coarse grainings of belief structures could have been finer or perhaps avoided altogether if an appropriate method were chosen. The important point for us is to suggest that this simulation paradigm has promise and should be useful for studying other quantum games. 

More speculatively, then, it may also suggest an avenue worth exploring in the study of quantum algorithm design. 
As an algorithm is a plan of action, one imagines a good algorithm should be rationally preferred by an agent given the right kinds of initial priors. Might rational agents playing the right kind of game stumble upon algorithms we haven't yet discovered? 

Our work also inspires a general study of agential sensitivity to resources. We have seen agents sharpening their beliefs in entanglement, but it would be interesting to explicitly probe other resources or to work backwards and see whether novel resources can actually be extracted from the structure of an experienced agent's beliefs. 

Our approach aligns with the basic ethos of epistemic game theory, where players' beliefs about opponents' beliefs are an essential ingredient in game analysis. From an analytic perspective, there has been work to extend theorems of epistemic game theory to the quantum domain, most notably focusing on Aumann's agreement theorem~\cite{Aumann1976,Khrennikov2015,Contreras-Tejada2021,Leifer2022}. Explicitly adopting the QBist perspective may suggest different approaches to pursuing these kinds of quantum analogs. 

From a simulation perspective, work in this area also suggests aspects which might be modified to interesting effect. For example, so-called type structures \cite{Perea2012,Dekel2015} are often used to represent a player's beliefs and to impose particular belief constraints beyond rationality instead of explicitly specifying a belief hierarchy. Similarly, it is sometimes of interest to consider the consequences of some players failing to meet the criterion of rationality~\cite{Dekel2015}. In such situations, one might allow for agents to revise their beliefs about others' rationality, perhaps providing useful criteria for judging if interactions are actually with another agent. As another example, if a probability zero event happens to a real agent, Bayesian updating is unavailable so their best course of action is simply to reconsider their priors. We imposed a probability floor to avoid this difficulty for our simulated agents, but it could be worth exploring one of the more sophisticated modeling approaches based on lexicographic or nonstandard probability which have appeared to treat this scenario~\cite{Brandenburger2014,Halpern2010}. 

We expect combining a QBist understanding of quantum theory with an epistemic understanding of games to continue to be productive into the future.

\section*{Acknowledgments}
JBD acknowledges support from the National Science Foundation Grant PHY-2116246. PJL acknowledges support from STAQ under award NSF PHY-1818914/232580. The authors also acknowledge several helpful comments and critiques by an anonymous referee on an earlier draft of this paper.

\bibliography{bibliography}

\end{document}